\documentclass[aps,prb,twocolumn,showpacs,amsmath,amssymb,superscriptaddress]{revtex4-2}
\usepackage{graphicx}
\usepackage{color}
\usepackage[colorlinks,bookmarks=false,citecolor=blue,linkcolor=red,urlcolor=blue]{hyperref}
\usepackage{ulem}
\usepackage{tabularx}
\usepackage[nounderscore]{syntax}

\newcommand{\be}{\begin{equation}}
\newcommand{\ee}{\end{equation}}

\begin{document}
\title{Surface critical behaviors of coupled Haldane chains  }
\author{Wenjing Zhu}
\affiliation{Department of Physics, Beijing Normal University, Beijing 100875, China} 
\author{Chengxiang Ding}
\affiliation{School of Science and Engineering of Mathematics and Physics, Anhui University of Technology, Maanshan, Anhui 243002, China}
\author{Long Zhang}
\affiliation{Kavli Institute for Theoretical Sciences and CAS Center for Excellence in Topological Quantum Computation, University of Chinese Academy of Sciences, Beijing 100190, China}
\author{Wenan Guo}
\email{waguo@bnu.edu.cn}
\affiliation{Department of Physics, Beijing Normal University, Beijing 100875, China} 
\affiliation{Beijing Computational Science Research Center, Beijing 100193, China}
	\date{\today}

\begin{abstract}
	The special surface transition at (2+1)-dimensional quantum critical point is precluded in corresponding classical critical point. The mechanism of such 
	behavior which is only found in dimerized Heisenberg models so far is still under debate. To illuminate the role of symmetry protected topological 
	(SPT) phase in inducing such nonordinary behaviors, we study a system on a two-dimensional square lattice consisted by interacting spin-1 Haldane chains, which
	has a genuine SPT phase--the Haldane phase--at weak interchain interactions and a quantum critical point belonging to the 
	classical 3D O(3) universality class to the N\'eel phase. Different from models studied previously, there is no dimerization in the current model. 
	Cutting the system along the chain direction or perpendicular to the chain direction 
	exposes two different surfaces. Using unbiased quantum Monte Carlo simulations,
	we find that the two different types of surface show completely different surface critical behaviors at the bulk critical point, 
	resulted from different surface states in the SPT phase. 
	For the system with surfaces along the chain direction, the surface critical behavior is of ordinary type of the bulk 3D O(3) critical point, while
	for the surfaces perpendicular to the chain direction, the surface critical behavior is nonordinary, consistent with special transitions found 
	in dimerized Heisenberg models.  Our numerical results demonstrate that 
	the gapless surface state in the gapped SPT phase together with the gapless mode of critical 
	point is a pure quantum scenario that leads to the nonordinary transition. 
	
\end{abstract}

\pacs{to be added }

\maketitle

\section{Introduction}
	\label{intro}

The field of classical surface criticality behavior (SCB) is rather mature.  \cite{binder, BH}
It is well known and understood well based on the renormalization group theory \cite{Diehl} that, by tuning the surface coupling, different 
surface universality classes can be realized. In particular, in the case that the surface 
can hold a long-range order, e.g., the two-dimensional (2D) surface of the three-dimensional (3D) Ising model, three different SCB universality 
classes: ordinary, extraordinary and special transitions, can be realized.  
But for models with continuous symmetry that can not hold a long-range order on the surface at finite temperature due to Mermin-Wagnar theorem, \cite{Mermin}  
e.g. the 2D surface of a 3D O(n) model with $n \ge 3$, it is widely accepted that there is neither extraordinary nor special transition.  
According to the quantum to classical mapping valid in particular to unfrustrated quantum antiferromagnets(AF),  the
$d$-dimensional SU(2) quantum critical point can be described by a ($d+1$)-dimensional classical effective O(3) Ginzburg-Landau
theory. \cite{Sachdev}  We thus expect that there is neither special nor extraordinary surface transition at the 2D SU(2) quantum critical points.

However, a nonordinary SCB was found in the 2D Affleck-Kennedy-Lieb-Tasaki (AKLT)\cite{AKLT} to N\'eel quantum critical point, which belongs to the 
3D O(3) universality class, recently. \cite{zhanglong}  The bulk quantum criticality is realized by the spin-$1/2$ AF Heisenberg (AFH) model on a decorated square lattice (DS). The 
surface transition has exponents different from those of the ordinary transition of the 3D O(3) critical point,
suggesting that an impossible SCB happens at the quantum critical point, denying the quantum to classical correspondence.
Such an unusual behavior was attributed to the symmetry-protected topological (SPT) \cite{GuWen, Pollmann} phase 
which, in contrast to the trivial (non-SPT) disordered phase, has gapless surface states. 
At quantum critical point, such pure quantum originated gapless states together with the gapless critical modes leads to unconventional SCB.\cite{zhanglong}

This discovery has inspired further interests in the investigation on SCBs with pure quantum origin. 
Ding et al \cite{Ding2018} found that different ways of cutting 2D periodic dimerized Heisenberg models on the square lattice into systems with boundaries 
can lead to all three types of SCB universality at the bulk quantum critical points which belong to the 3D O(3) universality class. 
In particular, for the columnar dimerized Heisenberg model, system with surfaces formed by non-dangling spins shows SCB corresponding to the ordinary
transition of the 3D O(3) class, while, remarkably, 
system with surfaces formed by dangling spins shows nonordinary SCB, with exponents in agreement with the special transition of the 3D O(3) class---
numerical values of the exponents are close to those obtained by the renormalization-group calculations for the special surface transition of the 3D O(3)
model. Similar results are obtained by Weber et al \cite{Weber1} independently. 

Furthermore, Weber and collaborators studied the SCB at the same quantum critical point between the AKLT phase 
and the N\'eel phase of the spin-1/2 Heisenberg model on the DS lattice studied in \onlinecite{zhanglong} but with different cut introduced. They found that 
the system shows the ordinary SCB on the non-dangling surface, although it shows
the same nonordinary transition as that found by Zhang and Wang \cite{zhanglong} on the dangling surface. This finding challenges the role of SPT in the 
origin of nonordinary SCBs.
In addition, they studied another quantum critical point separating the plaquette valence-bond crystal (PVBC) 
phase and the N\'eel phase of the same model. Nonordinary and ordinary transitions are again found at the dangling and nondangling surfaces, respectively. 

The exponents for the nonordinary transitions among various models agree well,
therefore they are taken as an indication for a distinct universality class: the special transition of the 3D O(3) class. 
Nevertheless it was also observed that the exponents varies upon perturbations, thus the universality is less universal than previously anticipated.\cite{Weber2}

Although it is apparent that the nonordinary SCBs have purely quantum origin, the mechanism is not clear yet. 
One possible scenario is that the dangling surface forms spin-1/2 AFH chain. Because of the topological $\theta$-term, the proliferation of topological 
defects in the corresponding classical field theory are suppressed and the surface captures the gapless state, complemented with the critical mode 
of the bulk, this leads to the special transition\cite{Ding2018,Weber1}. However, a recent work by Weber and Wessel\cite{Weber2} shows that this scenario is 
problematic: the dangling surface of spin-1 AF Heisenberg model also shows unexpected nonordinary surface transition. The role of SPT in the nonordinary SCB, 
which was challenged \cite{Ding2018,Weber1}, as mentioned in previous text, was further objected by the finding that
the suggested AKLT phase of the spin-1/2 Heisenberg model on the DS lattice can be adiabatically connected to the quantum-disorder state of the bilayer 
Heisenberg model on square lattice without breaking any symmetries, which suggests that the disordered phase is not a real SPT phase. \cite{Weber1}

\begin{figure}[!ht]
\includegraphics[width=1.0 \columnwidth]{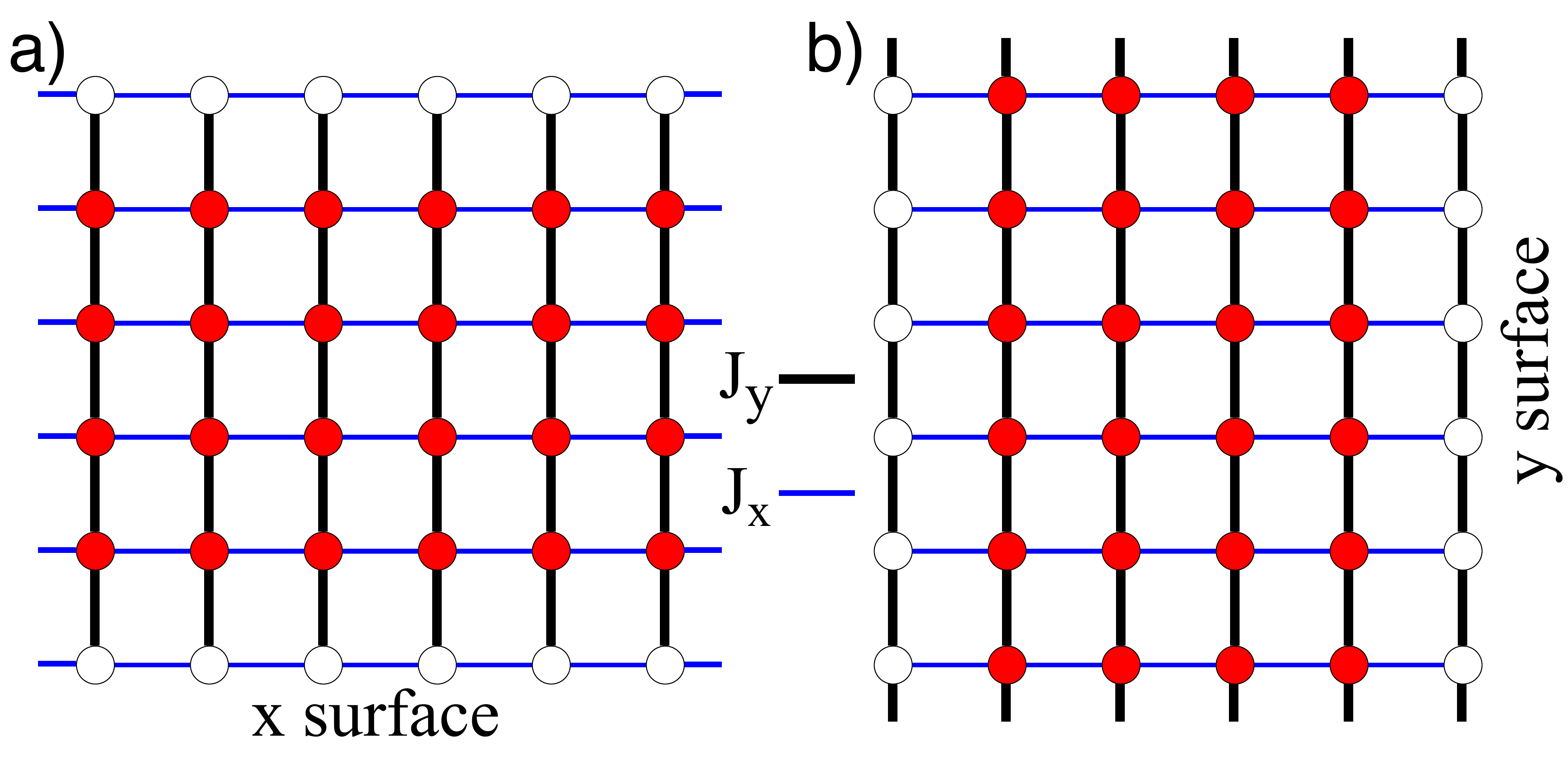}
\caption{ The coupled $S=1$ vertical Haldane chains form 2D square lattice with two different settings of boundaries. 
	(a) periodic boundary condition are applied in the x direction while open boundaries in the y direction. 
Two surfaces along the x direction are exposed. (b) periodic boundary condition are applied in the y direction while open boundaries are used in the x direction. 
	 Two y surfaces 
are formed.    $J_x$ (horizontal blue bonds ) and $J_y$ (vertical bold black bonds) are inter-chain and intra-chain couplings, respectively. 
\label{fig:lat}}
\end{figure}

In consideration of the current confusing situation, it is good to study a model with 
a genuine SPT phase separated from the N\'eel phase by a quantum critical point that belongs to 
the 3D O(3) universality class and to investigate if there is a nonordinary SCB related to the SPT gapless surface mode,
but has nothing to do with dimerization, which defines the dangling and non-dangling edges of spins.
In this work, we study coupled Haldane chains 
forming a 2D square lattice, with intrachain interaction $J_y$ and interchain interaction $J_x$, as shown in Fig. \ref{fig:lat}.  
The model describes materials that attracted considerable research 
\cite{Mutka,Asano,Honda,Renard1, Renard2, Regnault, Takigawa1, Takigawa2, Zaliznyak, Uchiyama,Pahari,Gadet,Darriet}. 

In one dimension (1D), the Haldane phase \cite{Haldane,Affleck} with the SPT order \cite{GuWen,Pollmann} is characterized by nonlocal string order \cite{denNijs}
parameter. With open boundaries, the Haldane chain carries gapless spin-1/2 
excitation. This is easy to understand from the AKLT state by deforming $S=1$ spin into two $S=1/2$ spins.
In 2D, the model has a gapped Haldane phase for $J_x \ll J_y$.   
Although the string order parameter is argued to decay exponentially for arbitrarily small interchain coupling\cite{Anfuso}, 
it is demonstrated that the Haldane phase remains a nontrivial SPT state for small but finite interchain coupling. \cite{WS3}
It is clear that the model can be adiabatically connected to the Haldane phase in 1D as $J_x \to 0$. As a result, the system should be able to present 
nontrivial surface states that are either gapless or degenerate edge states. The spatial inversion symmetry 
about the chain protects edge states of surfaces along the x direction against dimerization.\cite{WS3, Pollmann1}

Increasing interchain coupling $J_x$, the model will be brought into a N\'eel phase. The quantum critical point between the SPT Haldane and the N\'eel 
phase \cite{WS3,ME,Sakai,Koga,Kim,Matsumoto2001,WS1} locates at $J_x=0.043648(8)$ and the transition belongs to the 3D O(3) universality class. \cite{Matsumoto2001} 

To study the SCBs of the model, we use periodic boundary condition along one direction and make open boundary boundaries along the other direction by cutting 
a row/column of bonds perpendicular to this direction. The spins connected by these bonds form two surfaces. Figure \ref{fig:lat}(a) shows a system with 
periodic boundary in the x direction and open boundaries in the y direction. Two x surfaces are formed by cutting a row of $J_y$ bonds. Figure \ref{fig:lat}(b) 
shows a system with periodic boundary in the y direction and open boundaries in the x direction. In this case, two y surfaces are exposed. 
 
The model is suitable for checking the role of the symmetry-protected gapless surface state in inducing the special SCB, because,
different from previous works \cite{zhanglong, Weber1}, there's no dimerization, therefore, no dangling or non-dangling surface.  
Using unbiased quantum Monte Carlo (QMC) simulations, we find that, in the gapped Haldane phase, 
the string order parameter ${\cal S}$ of the chain, either with open ends or periodic, decays exponentially with system size $L$.  
This is different from simulation results reported in [\onlinecite{WS3}], but in agreement with the theoretical prediction\cite{Anfuso}.
The spin-spin correlation along the surface decays algebraically for the x surfaces, but exponentially for the y surfaces.
This is attributed to the fact that, in the SPT phase, the latter surfaces (y surfaces) are gapped, while the former surfaces are gapless, because 
a spin-1 model can has spin-1/2 excitations. 
Furthermore, we find out that the system with x surfaces 
shows special SCB, while the system with 
y surfaces 
shows ordinary SCB. 
This proves that the nontrivial surface states of a SPT state can induce nonordinary SCB.

The paper is organized as follows: In Sec. \ref{MM}, we introduce the model, discuss the bulk quantum phase transition and describe the quantum Monte
Carlo methods.   In Sec. \ref{spt}, we study the properties of the SPT phase. We investigate the string order parameter and its decaying behavior 
when couplings among chains are included. In particular, we discuss the string order parameter in the case that surfaces are presented. We also show the 
surfaces are gapless when the bulk are gapped in the SPT phase.  We then discuss the SCBs at bulk critical point in Sec. \ref{scb}.  In subsection \ref{ordinary},
we present our results on the ordinary transition along x surfaces.  In subsection \ref{special},  the results of special SCB are presented. 
We conclude in Sec. \ref{concl}.

\section{Models and method}
\label{MM}

The model we study here is $S=1$ AFH chains (Haldane chains) coupled by
weak interchain interactions, forming a 2D square lattice, as illustrated in Fig. \ref{fig:lat}.
The Hamiltonian of the model is given by
\begin{equation}
H = J_x \sum_{\langle i, j \rangle_x } {\bf S}_i  \cdot {\bf S}_j + J_y \sum_{\langle i, j \rangle_y } {\bf S}_i  \cdot {\bf S}_j ,
\end{equation}
where $\langle i, j \rangle_y$ 
are nearest neighbors along the chains with the coupling strength $J_y>0$, while $\langle i, j \rangle_x$  are nearest neighbors 
in two neighboring chains, with the interchains coupling $J_x>0$.
We set $J_y=1$ and consider $g\equiv J_x < J_y$. The system has a quantum critical point (QCP) separating the Haldane phase and the N\'eel phase at $g_c=(J_x)_c=0.043 648(8)$.\cite{Matsumoto2001}  
The transition  belongs to the 3D classical O(3) universality class with the correlation exponent $\nu$ estimated to be 0.70(1) and $\gamma=1.373(3)$. 
\cite{Matsumoto2001}

For $0<g<g_c$, the bulk is at a gapped disordered Haldane phase,  
which is a SPT phase \cite{WS3} with y-parallel Haldane chains behaving as independent $S=1$ AFH chains.
However, the string order parameter is zero, because the string order is unstable under the
perturbation of interchain couplings \cite{Anfuso}.
Cutting the lattice along the $x$ direction, 
each independent $S=1$ Haldane chain carries two effectively degenerate $S=1/2$ 
spins at the ends, which form the x surfaces of the system.  These effective spins  couple each other on the edge along the $x$ direction by weak
interchain couplings, forming two gapless $S=1/2$ AFH chains. 
On the other hand,  
cutting along the $y$ direction exposes two surfaces formed by two periodic $S=1$ Haldane chains, thus does not introduce gapless edge states. 
The two different cuts are graphed in Fig. \ref{fig:lat}.

In this work, we use the Stochastic Series Expansion (SSE) \cite{SSE} quantum Monte Carlo (QMC) 
method with the directed loop updating algorithm \cite{directedloop} to study the system, with special attention put on the SPT phase and the surface critical 
behavior induced by the SPT physics and bulk critical mode.  The simulations are performed on square lattices.  
To avoid large corrections to scaling due to the strong spatial anisotropy in the couplings
\cite{Matsumoto2001}, we set aspect ratio $R= L_y/L_x=4$ where $L_x$ and $L_y$ is the linear size in the x and y direction, respectively. Henceforth, we use 
$L=L_y$ as the linear size of the system. The inverse 
temperature $\beta$ is scaled as $L$ to probe ground state properties, considering the dynamical critical exponent $z=1$ for the bulk criticality.  

\section{Symmetry protected topological phase}
\label{spt}  

In 1D, the Haldane phase is characterized by the hidden nonlocal string order \cite{denNijs}, which is defined as follows
\begin{equation}
\mathcal{S}(i, j) = \langle S^z_i \exp( i \pi \sum_{k=i+1}^{j-1}S^z_k)S^z_{j}\rangle
	\label{SO}
\end{equation}
where $i, j$ are two spins in the chain and $k$ labels a spin in between of them. The string order is finite for $|i-j| \to \infty$ in the 1D Haldane phase.  
This definition can also be used for the 2D spin-1 AFH system by restricting $i,j$ and $k$ as spins along an individual chain. 

It was predicted that the string order is not stable and decays exponentially for arbitrarily weak interchain 
coupling by means of perturbation theory and bosonization \cite{Anfuso}. However, it was shown numerically that the string order parameter 
decays algebraically in the Haldane phase of coupled Haldane chains according to finite-size data up to linear size $L=72$ at interchain couplings much weaker than critical 
value. \cite{WS3} 
Therefore it is necessary to investigate the string order parameter for various values of $g<g_c$ in the Haldane phase, 
including those that are close to the critical point. We will calculate the parameter up to much larger system sizes than those reached in the literature,
so that we can pin down the scaling behavior of the string order parameter in the 2D Haldane phase. 
Besides, we will investigate the string order parameter in different boundary conditions. 

\begin{figure}
\includegraphics[width=\columnwidth]{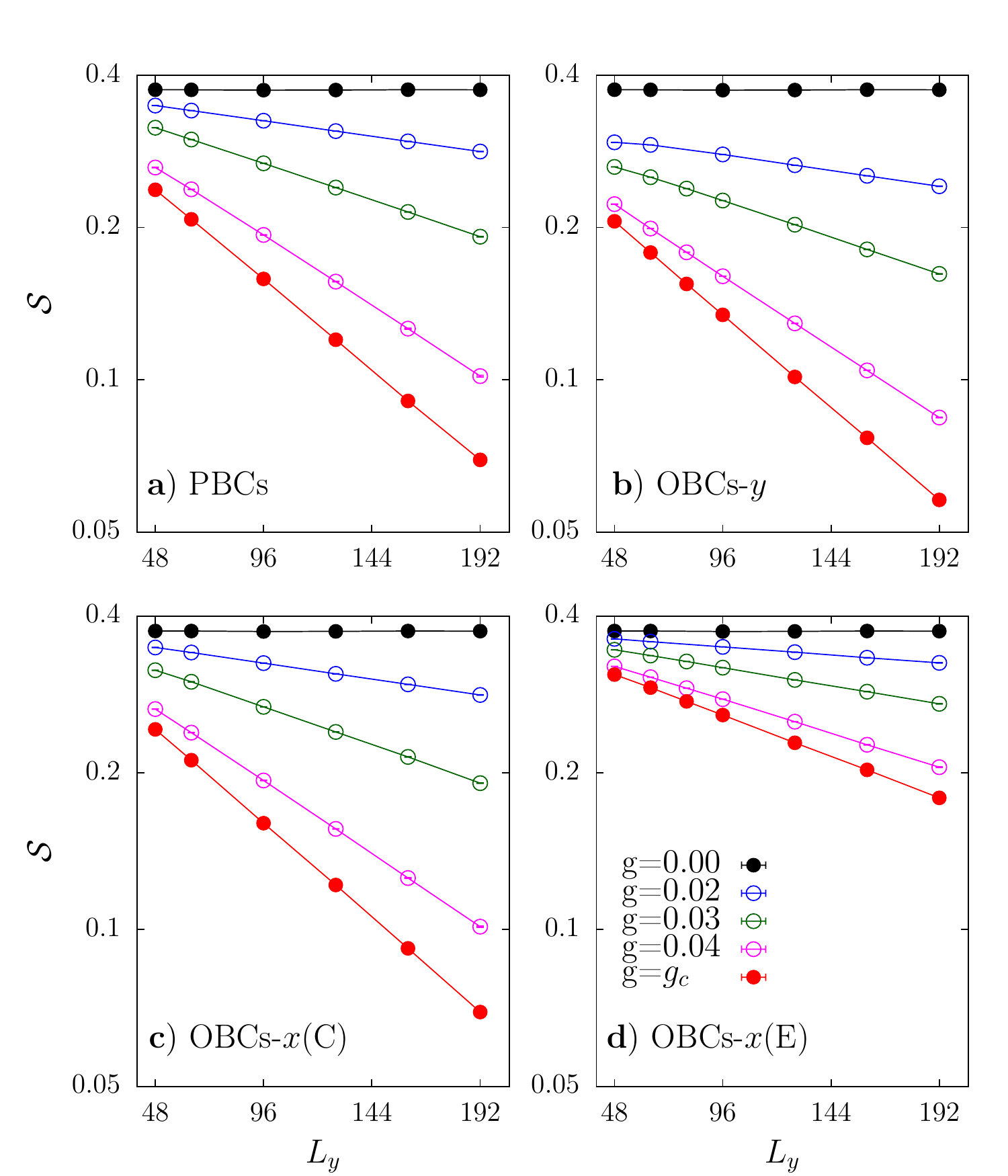}
\caption{String order parameter of a chain at the maximum available distance ${\cal S}(L/2)$ at different interchain couplings for systems 
with (a) periodic boundaries, (b) open boundaries in the y direction, 
	and (c)(d) open boundaries in the x direction, shown on a linear-log scale.
	For (b), the chain has open ends at the x surfaces where the starting spin locates.  For (c) the chain is at the center of the system,  
	and, for (d), the chain is one of the y surfaces.   }
\label{fig:String}
\end{figure}

We focus on the string order parameter ${\cal S}(L/2)$ at the maximum available distance $|i-j|=L/2$ of a chain along y direction.
The results are shown in Fig. \ref{fig:String}. 

For systems with periodic boundary conditions (PBCs) in both x and y directions, the results are shown in Fig. \ref{fig:String}(a). 
We find that, for $g<g_c$, the nonlocal string order decays exponentially with system size as it is 
large enough, in agreement with the theoretical prediction of \onlinecite{Anfuso}. For small $g$, e.g., $g=0.01$,  which is studied in \onlinecite{WS3}, such 
exponentially decaying behavior are clearly seen only when system sizes are much larger than 72. 

We then study the nonlocal string order parameter for  systems with periodic boundary condition along one direction, but open boundary condition along
the other direction with two surfaces exposed, as shown in Fig.\ref{fig:lat}. 
For the configuration Fig. \ref{fig:lat}(a), we calculate $S(L/2)$ along a chain connecting the two x surfaces. The starting site in Eq. (\ref{SO}) is chosen to 
be at the surface and the ending site sits at the center of the chain. 
The results are shown in Fig. \ref{fig:String}(b). The nonlocal string order parameter is found decaying exponentially with system size $L$.  
For the configuration Fig. \ref{fig:lat}(b), we study two situations. First, we calculate the string order parameter ${\cal S}(L/2)$ of the periodic chain forming the y surface.
The results 
are shown in Fig. \ref{fig:String}(d); Then we calculate ${\cal S}(L/2)$ of the chain sitting at the middle of the system. The 
results are shown in Fig. \ref{fig:String}(c). 
Again, we see exponential decay of the string order parameter.
Apparently, the open boundaries do not affect much the finite-size behavior of the nonlocal string order, which remains fragile as in periodic systems. 
 
\begin{figure}[h]
\includegraphics[width=1 \columnwidth]{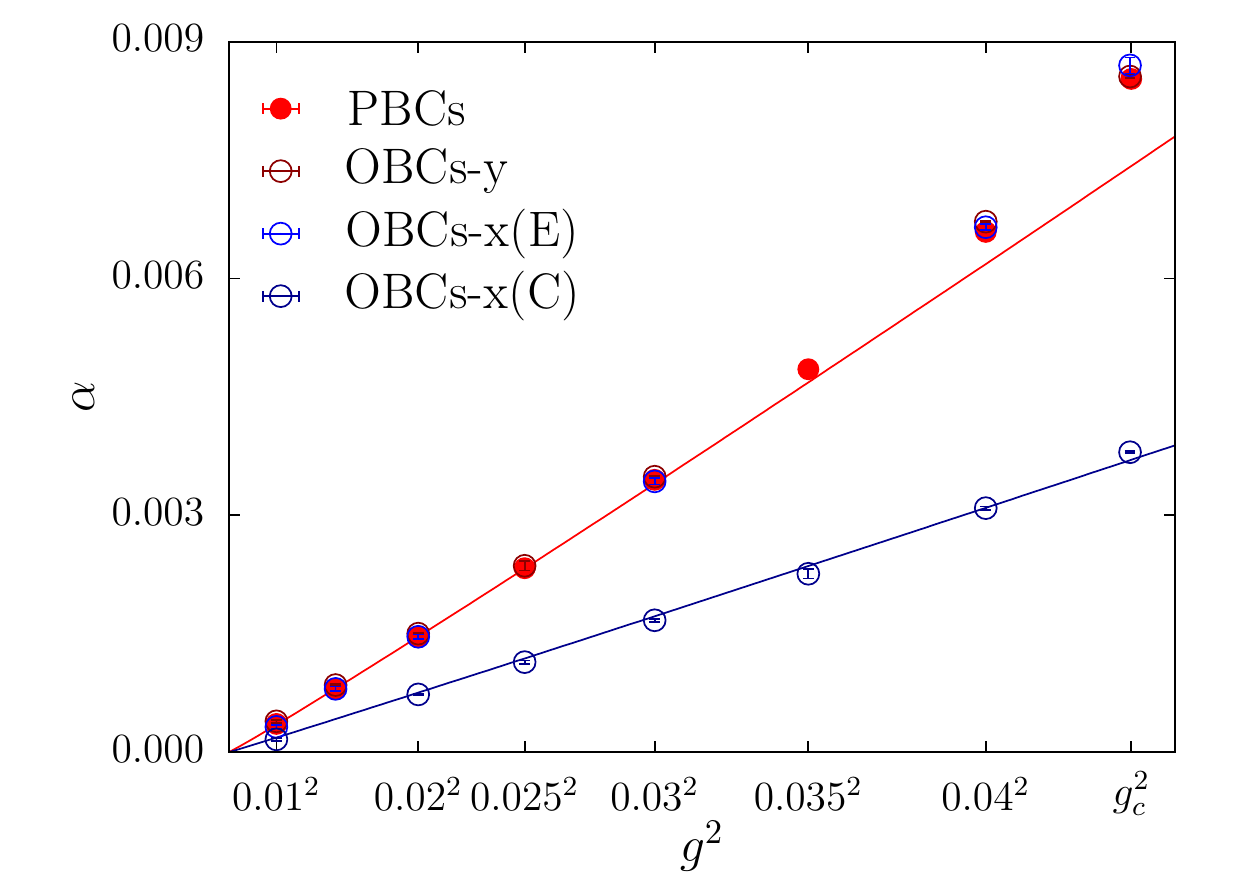}
	\caption{ 
	The inverse characteristic length $\alpha$ versus $g^2$ for systems with different boundary conditions.
	The two straight lines correspond to the scaling of the form  $\alpha(g)\propto  g^2$.
}
\label{fig:String_alpha}
\end{figure}

The exponentially decaying of the string order parameter is expected to behave in the following way 
\begin{equation}
 {\cal S}(L/2) \propto \exp{(-\alpha L/2)},
 \label{strfss}
 \end{equation}
with $\alpha\sim g^2$ in the case of $g$ being small enough.\cite{Anfuso}
Similar to correlation functions, \(\alpha\) could be regarded as the inverse characteristic length. 
We fit ${\cal S}(L/2)$ of systems with all boundary conditions at several values of $g$ below \(g_c\) to Eq.(\ref{strfss}).
The fits are carried out with data up to system size $L=192$. To obtain statistically sound fits,  small systems with $L<48$ are excluded.
The obtained $\alpha$ as functions of $g^2$ for different boundary conditions are graphed in
Fig. \ref{fig:String_alpha}.
We then fit the estimated $\alpha(g)$ for various boundary conditions
to the function $\alpha = c g^2$ with $c$ an unknown parameter. 
We find perfectly consistency to the theory for \( g \le 0.03 \), 
as illustrated in Fig. \ref{fig:String_alpha}.

Nonetheless the string order parameter vanishes in 2D, the Haldane phase remains a gapped weak SPT phase\cite{WS3},
which is fabricated by alignment of 1D SPT state. 
The 1D Haldane chain with open ends supports degenerate $S=1/2$ edge states. 
In 2D, we expect the open ends forming two surfaces which can be considered as two gapless $S=1/2$ AFH chains, 
as shown in fig. \ref{fig:lat}(a), while the bulk ground state is gapped. 
These gapless states are protected by the translational symmetry along the x direction.\cite{3DSPT}

To show the above expectations are correct, we study the equal-time spin-spin correlation defined as 
\begin{equation}
	C(i,j)=\langle (S_i^x S_j^x+S_i^yS_j^y)\rangle
\end{equation}
that equals to $2 \langle S_i^z S_j^z \rangle$ by symmetry and can be calculated through 
the following Green function
\begin{equation}
	G(i,j)=\frac{1}{2} (\langle S_i^+ S_j^-\rangle +\langle S_i^-S_j^+\rangle),
	\label{Green}
\end{equation}
which can be measured efficiently with improved estimator by using the loop updating algorithm in the SSE QMC simulations.\cite{Evertz1}

We focus on two specific correlations.  One is $C_\parallel(L_\alpha/2)$ that averages $C(i,j)$ between two surface spins $i$ and $j$ at a distance $L_\alpha/2$
over the $\alpha=x$ or $y$ surface under consideration.
The other is $C_\perp(L_\alpha/2)$ that 
averages $C(i,j)$ between two spins $i$ and $j$ at a distance $L_\alpha/2$ with $i$ fixed on the surface and the other spin $j$ 
located at the center of the bulk, with the direction $i$ to $j$  perpendicular to the surface, along the $\alpha$ direction.
  
\begin{figure}[h]
\includegraphics[width=1 \columnwidth]{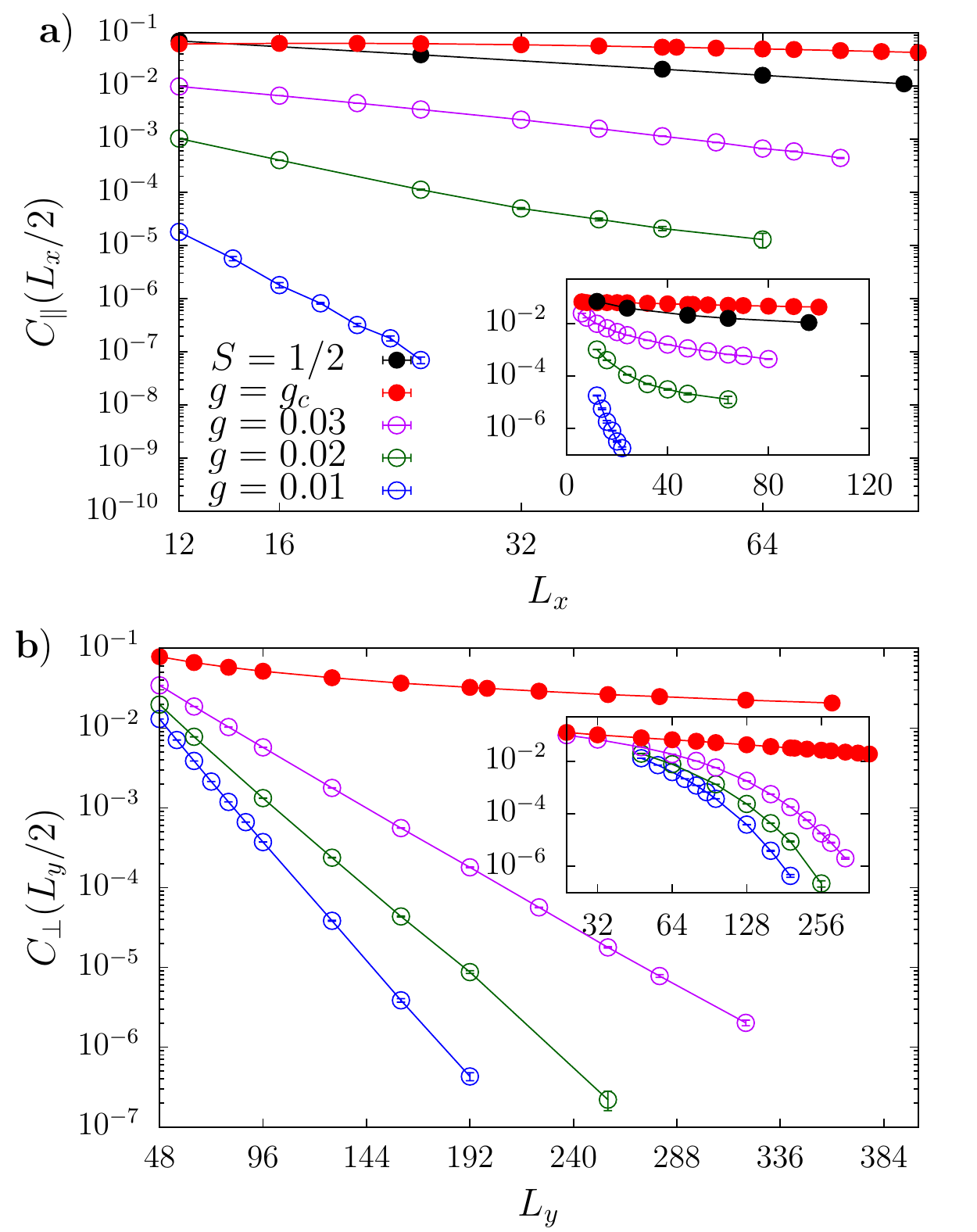}
\caption{
	Correlation functions $C_\parallel(L_x/2)$  (a) and $C_\perp(L_y/2)$ (b) of the systems with x surfaces,
	as illustrated by Fig. \ref{fig:lat}(a), at the SPT phase with different $g$ up to $g_c$.   
        For the purpose of comparison, $C_\parallel(L/2)$ for the $S=1/2$ AFH chain is also shown here.  
	(a) The main plot is set on a log-log scale, the 
	inset is on a linear-log scale. (b) The main plot is on a linear-log scale and the inset is on a log-log scale.}
 \label{fig:corr_xsurf}
\end{figure}

We first study the system with x surfaces. 
The results of $C_\parallel(L_x/2)$ at several $g \le g_c$ as functions of system size $L_x$ are plotted on a log-log scale in 
Fig. \ref{fig:corr_xsurf}(a). We find that $C_\parallel (L_x/2) $ decays algebraically with (large enough) system size $L$ for all $g$, including the critical $g_c$, indicating 
that the surface state is gapless, or critical. This is further manifested by the inset in which the correlations are plotted on a linear-log scale.  

More precisely, we expect the following finite-size scaling ansatz for $C_\parallel$
\begin{equation}
C_\parallel (L/2) = L^{-(d+z-2+\eta_\parallel(g))}( a + b L^{-\omega}), 
\label{eqn:scal1}
\end{equation}
where $\eta_\parallel(g)$ is the surface anomalous dimensions, which depends on $g$. 
$\omega>0$ is the effective exponents controlling  corrections to scaling. $d+z$ is the space-time dimension.  In the present model, $d=2$ and $z=1$. 

The largest system size we reached is $L_y=320$ and $L_x=80$.  
By excluding small system sizes gradually, we can obtain statistically sound fits of Eq.(\ref{eqn:scal1}) for all $g$,
with the correction to scaling term removed. For example, for the case $g=0.03$, we find $1+\eta_\parallel=1.82(3)$ 
by fitting Eq.(\ref{eqn:scal1}) with the correction to scaling term discarded to the finite-size data in the range of \(48 \leq L_x \leq 80 \). 
The reduced \(\chi^2\) of the fit is 1.58 with the goodness of fit defined as the P-value of the $\chi^2$ distribution $P=0.175$. 
For $g=0.02$, we have to exclude sizes smaller than $L_x=32$ to obtain satisfying fit.  
We have also tried to include multiplicative logarithmic correction to the simple power law decay as in the 1D spin-1/2 AFH chain, but no improving was found 
in the fitting, and the decay exponents $1+\eta_\parallel$ are always close to 2. 
We conclude that, although the edge is gapless, the anomalous exponent is not consistent with the $1/r$ decay with multiplicative logarithmic corrections for the 
spin-1/2 AFH chain. \cite{AGSZ, SFS, GS} 
For comparison, $C_\parallel(L/2)$ for the $S=1/2$ AFH chain is shown in Fig. \ref{fig:corr_xsurf}(a). 
It is evident that the x surface can not be considered as a spin-1/2 AFH chain naively. 

Meanwhile, as shown in Fig. \ref{fig:corr_xsurf}(b) on a linear-log scale,  $C_\perp(L_y/2)$ 
exponentially decays with system size $L_y$ at all $g < g_c$, indicating that the bulk is gapped for $g<g_c$. 
This is further exhibited by the log-log plot in the inset. For $g=0.03$, a fit to the function $C_\perp(x)=a \exp(-x/\xi)$ with 
finite-size data in the range $128 \le L_y \le 320$ finds statistical sound results $\xi=27.8(1)$, with reduced $\chi^2=1.25$ and the P-value $P=0.28$.

\begin{figure}[h]
\includegraphics[width=1 \columnwidth]{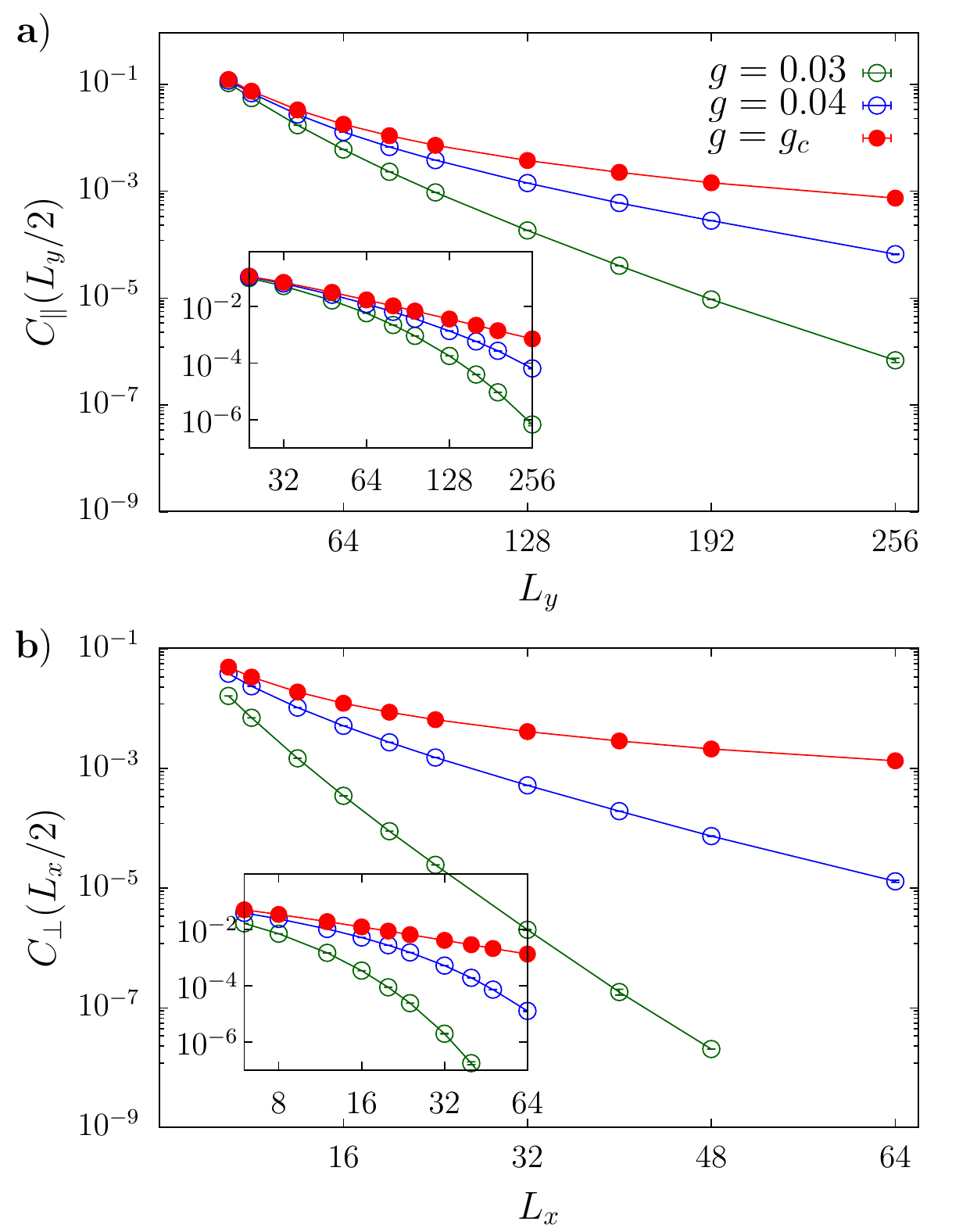}
\caption{For system with surfaces shown in Fig. \ref{fig:lat}(b) at the SPT phase with different $g$ up to $g_c$, (a) correlation along the surface 
$C_\parallel(L_y/2)$ vs $L_y$, (b) correlation perpendicular to the 
	surface $C_\perp(L_x/2) $ vs $L_x$. The two main plots are set on linear-log scales and the two insets are on log-log scales. }
\label{fig:corr_ysurf}
\end{figure}

On the other hand, for the system with open surfaces along y
direction, exposed by cutting a column of  interchain $J_x$ bonds, the surfaces are two periodic spin-1 chains weakly coupled to the bulk, as 
illustrated in Fig. \ref{fig:lat}(b). We find that the surface states is gapped, even though the string order parameter vanishes with finite 
interchain coupling included. 
This is obtained from the results of $C_\parallel(L_y/2)$ at several $g$ as functions of system size $L_y$, as shown in 
Fig. \ref{fig:corr_ysurf}(a).  
Although there are strong finite-size effects, we can fit 
$C_\parallel (L_y/2) $ to exponentially decaying forms of system size $L_y$ for large enough system sizes,  which indicates the surface 
states are gapped at the Haldane phase. 
Meanwhile, as shown in Fig.\ref{fig:corr_ysurf}(b),  $C_\perp(L_x/2)$ also 
exponentially decay with system size $L_x$ at $g < g_c$, reflecting the fact that the bulk is gapped for $g<g_c$. 

 \section{Surface critical behaviors}
 \label{scb}

 At bulk critical point,  some classical systems can have ordered surface if the interactions at the surface are enhanced. 
 When this happens, the ordered surface exhibits additional singular behavior, which is called as an extraordinary transition. 
 Furthermore fine-tune the surface couplings can lead the surface to a multicritical point, at which surface and bulk 
 are critical simultaneously. This is the special transition.  
 As discussed in Sec. \ref{intro}, according to quantum-classical corresponding, we expect no special and/or extraordinary surface
 transition in the present model, because there can not be a long-range order that breaks the continuous symmetry in 
 2D surface of a classical model at finite temperature due to Mermin-Wagnar theorem.\cite{Mermin} 
Here in this section, we show that, due to pure quantum effect, special surface transition can be realized at the surface perpendicular 
to the coupled Haldane chains without fine-tune surface couplings, while at another surface, the system shows the ordinary SCB. 

To study the surface critical behavior, we here, besides the correlations $C_\parallel$ and $C_\perp$, introduce  
 the squared staggered surface magnetization defined as 
\begin{equation}
	\begin{split}
		m^2_{s1}&= \frac{1}{2L_\alpha^2} [\langle (\sum_{i \in {\rm surface}} (-1)^{i} S^x_i)^2 \rangle + \langle (\sum_{i \in {\rm surface}} (-1)^{i} S^y_i)^2 \rangle],
	\end{split}
\end{equation} 
with $i$ labeling spins on the surface, which equals to $\frac{1}{L_\alpha^2} \langle (\sum_{i \in {\rm surface}} (-1)^{i} S^z_i)^2 \rangle$ by symmetry.
$m^2_{s1}$ can be expressed with the Green function defined in Eq.(\ref{Green})
\begin{equation}
	m^2_{s1}= \frac{1}{L_\alpha^2}\sum_{i,j \in {\rm surface}}(-1)^{i+j} G(i,j),
\end{equation}
with $i, j$ the spins on the surface and $\alpha=x$ for the surface along x-direction and $\alpha=y$ for the  y surface, respectively.

At bulk QCP, the surface should show power law surface critical behaviors.  Besides the Eq. (\ref{eqn:scal1}) with $\eta_\parallel=\eta_\parallel(g_c)$,
$m_{s1}^2(L)$ and $C_\perp(L)$ obey the  following finite-size scaling  forms \cite{BH}
\begin{equation}
	m_{s1}^2 L = c + L^{2y_{_{h1}}-(d+z)}(a + b L^{-\omega}),
\label{eqn:scal2}
\end{equation}
and
\begin{equation}
	C_\perp(L) = L^{-(d+z-2+\eta_\perp)}( a + b L^{-\omega}) .
\label{eqn:scal3}
\end{equation}
in which $y_{_{h1}}$ is the scaling dimension of the surface staggered magnetic field $h_1$, 
$\eta_\parallel$ and $\eta_\perp$ are two surface anomalous dimensions. 
The constant $c$ in Eq. (\ref{eqn:scal2}) results from the short-range nonuniversal part of $m_{s1}^2$. 
$\omega>0$ is the effective exponents  controlling  corrections to scaling.  
$d+z$ is the space-time dimension.  In the present model, $d=2$ and $z=1$. 

The surface critical exponents $y_{h1}$ , $\eta_\parallel$ and $\eta_\perp$ are expected to satisfy the following scaling relations\cite{Barber, Diehl, Lubensky}
\begin{equation}
2\eta_\perp = \eta_\parallel +\eta  
\label{eqn:relation1} 
\end{equation}
and
\begin{equation}
\eta_\parallel = d +z - 2y_{h1},
\label{eqn:relation2}
\end{equation} 
with $\eta$ the anomalous magnetic scaling dimension of the bulk transition in the $d+z$ spacetime. For our model, $d+z=3$ and 
 $\eta=0.0357(13)$ \cite{CHPRV}  is  the anomalous magnetic scaling dimension for the 3D O(3) universality class.

We now proceed to study systems with surfaces at its bulk quantum critical point to explore the surface critical behaviors.

\subsection{Ordinary transition}
\label{ordinary}
\begin{figure}[!h]
\includegraphics[width=1 \columnwidth]{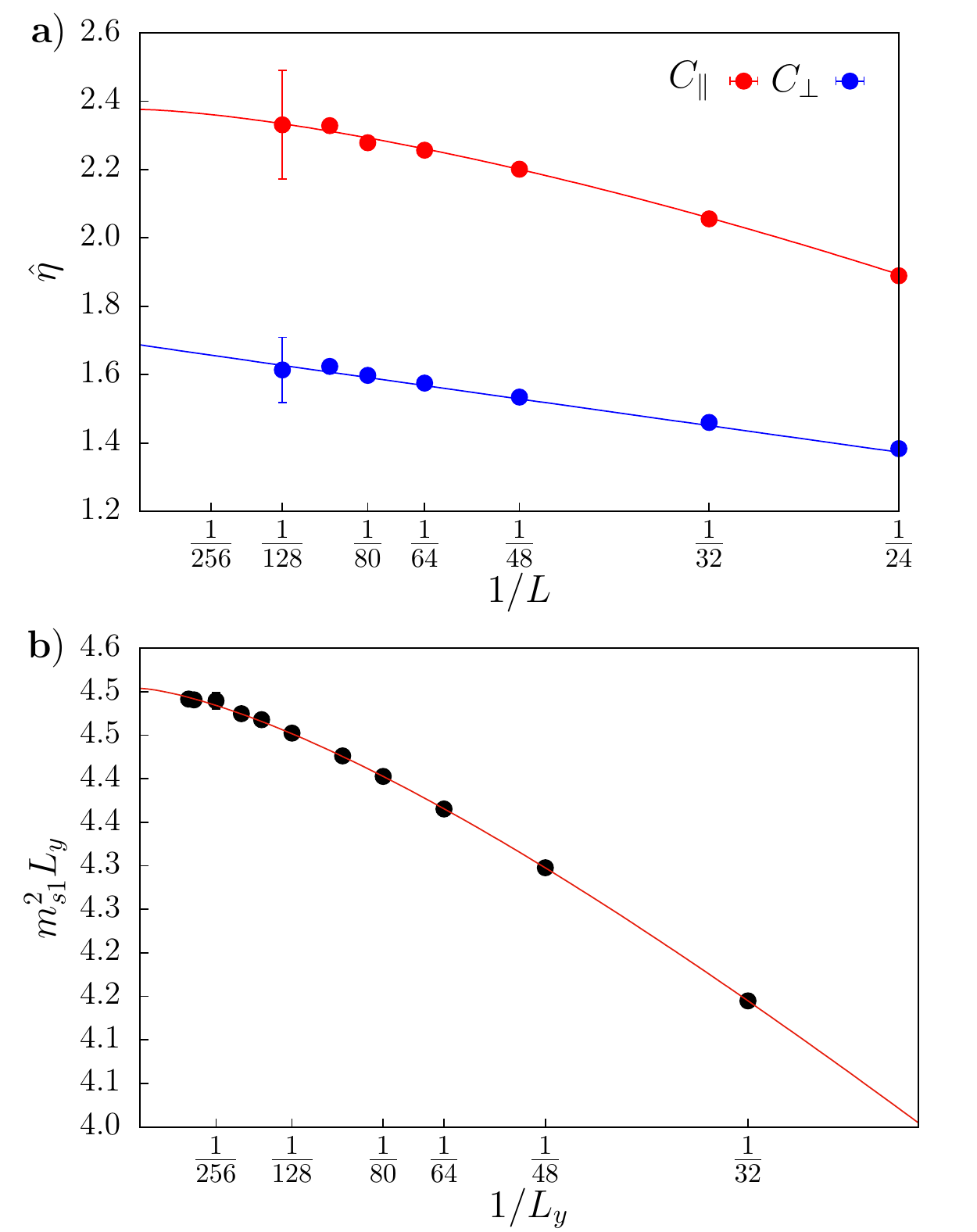}
\caption{ 
	Ordinary surface transition of the system configuration Fig. \ref{fig:lat}(b).
	(a) Finite-size dependence of effective surface anonymous exponents $\hat{\eta}(L)$
	for $C_\perp$(blue symbols)  and $C_\parallel$ (red symbols). The lines show fits giving the 
	estimates of $\eta_\perp$ and $\eta_\parallel$, respectively.  (b) 
	The rescaled squared staggered magnetization of the surface spins $m_{s1}^2 L_y$. The curve is a fit to the expected power law decay
	with a constant included.
\label{fig:S1Ac1gc}}
\end{figure}

We first consider the system that is cut to expose two y surfaces along the y direction, see Fig. \ref{fig:lat}(b). 
The surfaces are periodic spin-1 chains with length $L_y$.
To study the SCB of such a system at the bulk quantum critical point, 
the staggered surface magnetization  $m_{s1}^2$, spin correlations $C_\parallel(L_y/2)$ and $C_\perp(L_x/2)$ are calculated 
at $g_c$. The numerical results of $C_\parallel$ and $C_\perp$ are graphed in Fig. \ref{fig:corr_ysurf} and the results of rescaled $m_{s1}^2$ 
are plotted in Fig. \ref{fig:S1Ac1gc}(b). 

The spin correlations $C_\parallel(L_y/2) $ and $C_\perp(L_x/2)$ decay algebraically, as can be seen in the insets of Fig. \ref{fig:corr_ysurf}. This is 
different from the behaviors in the SPT Haldane phase, where both of them decay exponentially. 
We fit the data to finite-size scaling Eq. (\ref{eqn:scal1}) and (\ref{eqn:scal3}) to  find the anomalous surface scaling 
dimensions $\eta_\parallel$ and $\eta_\perp$. 

For large enough systems, the correction to scaling terms in Eq. (\ref{eqn:scal1}) and (\ref{eqn:scal3}) can be neglected. 
To estimate $\eta_\parallel$, we found the fit of $C_\parallel(L_y/2)$ is statistically sound for system sizes $L_y \ge 128$ and obtained 
$\eta_\parallel = 1.38(2)$. 
Using data of system sizes $L_x \ge 48$, we found statistically sound fit of $C_\perp(L_x/2)$ and obtained 
$\eta_\perp = 0.68(1)$.

To fit for more system sizes, we need to include the correction to scaling terms in Eq.(\ref{eqn:scal1}) and (\ref{eqn:scal3}). 
However, it's hard to estimate the value of the effective exponent $\omega$.
To verify our fits of $\eta_\perp$ and $\eta_\parallel$, we define an effective exponent $\hat{\eta}$ for a pair of systems with 
sizes $L$ and $2L$,
\begin{equation}
	\hat{\eta} (L)=\frac{1}{\ln(2)}(\ln C(2 L)-\ln C(L)),
	\label{eqn:etahat}
\end{equation}
where $C(L)$ stands for one of the two correlations $C_\parallel(L_y)$ or $C_\perp(L_x)$.
According to Eq.(\ref{eqn:scal1}) and (\ref{eqn:scal3}), $\hat{\eta}(L)$ should converge to $1+\eta$ in the following way
\begin{equation}
	\hat{\eta}(L)=1+\eta + c L^{-\omega},
	\label{eqn:zeff}
\end{equation}
where $\eta$ stands for $\eta_\parallel$ or $\eta_\perp$ for corresponding $\hat{\eta}$. 

As shown in Fig.~\ref{fig:S1Ac1gc}(a),
the two effective exponents converge to the $1+\eta_\parallel$ and $1+\eta_\perp$, respectively.
Fitting 
the data to Eq. (\ref{eqn:zeff}), we obtained statistically sound fits for data of sizes $L \ge 24$. 
The final estimations are $\eta_\parallel=1.378(9)$ with $\omega=1.45(4)$ 
and $\eta_\perp=0.69(2)$ with $\omega=0.99(5)$. 
The results are in good agreement with 
the estimations of $\eta_\parallel$ and $\eta_\perp$ obtained above. The final 
estimations are listed in Tab. \ref{tab_exp}. 

The rescaled surface magnetization $m_{s1}^2 L_y$  as function of system size $L_y$ exhibits a power-law decay to a nonzero 
constant $c$,  in the way of Eq. (\ref{eqn:scal2}). 
For large enough system sizes, the correction to scaling with effective exponent $\omega$ in Eq. (\ref{eqn:scal2}) can be dropped. 
A nonlinear fit based on Eq. (\ref{eqn:scal2}) finds  $c=4.552(1)$, suggesting that the transition 
is of ordinary type. The scaling dimension of the surface magnetic field is obtained as $y_{h1}=0.78(1)$. 
In the fit, we restricted the system sizes to $L_y \ge 64$ and, thus, neglected the correction to scaling term.
We have also tried to set $\omega=1.5$ in Eq. (\ref{eqn:scal2}) and fitted it to data of sizes $L_y \ge 24$. 
The fit has good reduced $\chi^2=1.18$ and p-value $P=0.31$, and finds $c=4.554(1)$ and 
$y_{h1}=0.79(1)$, which are consistent well with the fit without correction to scaling term but only for large enough system sizes.
The final estimate of $y_{h1}$ is listed in Tab. \ref{tab_exp}.

The exponents found here obey the scaling relations Eqs. (\ref{eqn:relation1}) and (\ref{eqn:relation2}). 
They also agree well with the corresponding exponents found for the ordinary surface transition of O(3) bulk criticality in other 2D quantum 
models \cite{zhanglong, Ding2018, Weber1, Weber2},
and are consistent with those of the ordinary transition of the 3D classical O(3) universality class \cite{Deng}. 

This is understandable, because we know that, for
the ordinary SCBs, the algebraic correlations are all bulk-induced behaviors.
This is true also for quantum critical system so far discussed.
However, all those ordinary SCBs found thus far in the quantum critical points of dimerized systems 
happen at surfaces formed by so called {\it non-dangling} spins.
But here, the interesting thing is that the y surface of the current model are weakly coupled to the bulk, therefore, are 
actually formed by kind of {\it dangling} spins.

\begin{ruledtabular}
  \begin{table*}[!t]
  \caption{ Results of Surface critical exponents of the coupled Haldane chains(CHC). For the convenience of the readers, results of the SCBs on other models with transitions in the 
	  3D O(3) universality class are also listed for comparison, with
	  CD-DAF the Dimer-AF QCP of the columnar dimerized Heisenberg model, SD-DAF the Dimer-AF QCP of the staggered dimerized Heisenberg models,
	  DS-DAF the Dimer-AF QCP of the dimerized Heisenberg model on the DS lattice; DS-PAF the PVBC-AF QCP of dimerized Heisenberg model on the DS lattice, 
	  3D CH  the three-dimensional classical Heisenberg model. For the types of cuts, D means dangling and N means nondangling. 
	  The field theoretical results(FT) from various methods are also listed for comparison. 
  }
  \label{tab_exp}
  \begin{tabular}{c|l|l|l|c|c|c} 
SCB class               &  Model/methods         &cuts        &     Spin S         &$\eta_\parallel$       &  $\eta_\perp$   & $y_{h1}$ \\
  \hline
Sp.                     & CHC                    &x surface   &     1              & -0.57(2)              &   -0.27(2)      &  1.760(3)   \\
Ord.                    &                        &y surface   &     1              & 1.38(2)               &    0.69(2)      &  0.79(2)   \\ 
  \hline
Sp. \cite{Ding2018}     & CD-DAF                 &  D         &     1/2            &  -0.445(15)           &    -0.218(8)    &1.7339(12)\\
Sp.  \cite{Weber1}      &                        &  D         &     1/2            &  -0.50(6)             &    -0.27(1)     &1.740(4) \\
Sp.   \cite{Weber2}     &                        &  D         &     1              &  -0.539(6)            &    -0.25(1)     & 1.762(3) \\
Ord.  \cite{Weber1}     &                        &  N         &     1/2            &  1.30(2)              &    0.69(4)      & 0.84(1)    \\
Ord.  \cite{Ding2018}   &                        &  N         &     1/2            &  1.387(4)             &     0.67(6)     & 0.840(17) \\   
Ord.  \cite{Weber2}     &                        &  N         &     1              &  1.32(2)              &    0.70(2)      & 0.80(1)    \\
\hline
Ord.  \cite{Ding2018}   &SD-DAF                  &  N         &      1/2           & 1.340(21)             &    0.682(2)     & 0.830(11) \\
     \hline
Sp.  \cite{zhanglong}   & DS-DAF                 & D         &      1/2           &   -0.449(5)           &   -0.2090(15)   & 1.7276(14) \\
Sp.  \cite{Weber1}      &                        & D         &      1/2           &   -0.50(1)            &   -0.228(5)     & 1.728(2)    \\
Ord.  \cite{Weber1}     &                        & N         &      1/2           &   1.29(6)             &    0.65(3)      & 0.832(8)    \\
\hline
Ord.  \cite{zhanglong}  & DS-PAF                  & N         &      1/2           &   1.327(25)           &    0.680(8)     & 0.810(20)\\
Ord.  \cite{Weber1}     &                         & N         &      1/2           &   1.33(4)             &    0.65(2)      & 0.82(2)    \\
Sp.   \cite{Weber1}     &                         & D         &      1/2           &  -0.517(4)            &   -0.252(5)     &  1.742(1)  \\
          \hline
Ord.  \cite{Deng}       &3D CH                    &           &      class.        &                       &                 & 0.813(2)\\

\hline
Ord.\cite{DD1}          & FT, 4-d $\epsilon$-exp  &           & class.             & 1.307                 &  0.664          & 0.846     \\
Ord.\cite{DN}           & FT, d-2 $\epsilon$-exp  &           & class.             & 1.39(2)               &                 &               \\
Ord.\cite{DS1,DS2}      & FT, Massive field       &           & class.             &  1.338                & 0.685           & 0.831      \\    
Ord.\cite{GLMR}         & FT, Conformal bootstrap &           & class.             &                       &                 & 0.831      \\           
Sp.\cite{DD2}           & FT, 4-d $\epsilon$-exp  &           & class.             &  -0.445               & -0.212          &  1.723    \\     
  \end{tabular}
  \end{table*}
\end{ruledtabular}

\subsection{Special transition}
\label{special}

\begin{figure}[!h]
\includegraphics[width=1 \columnwidth]{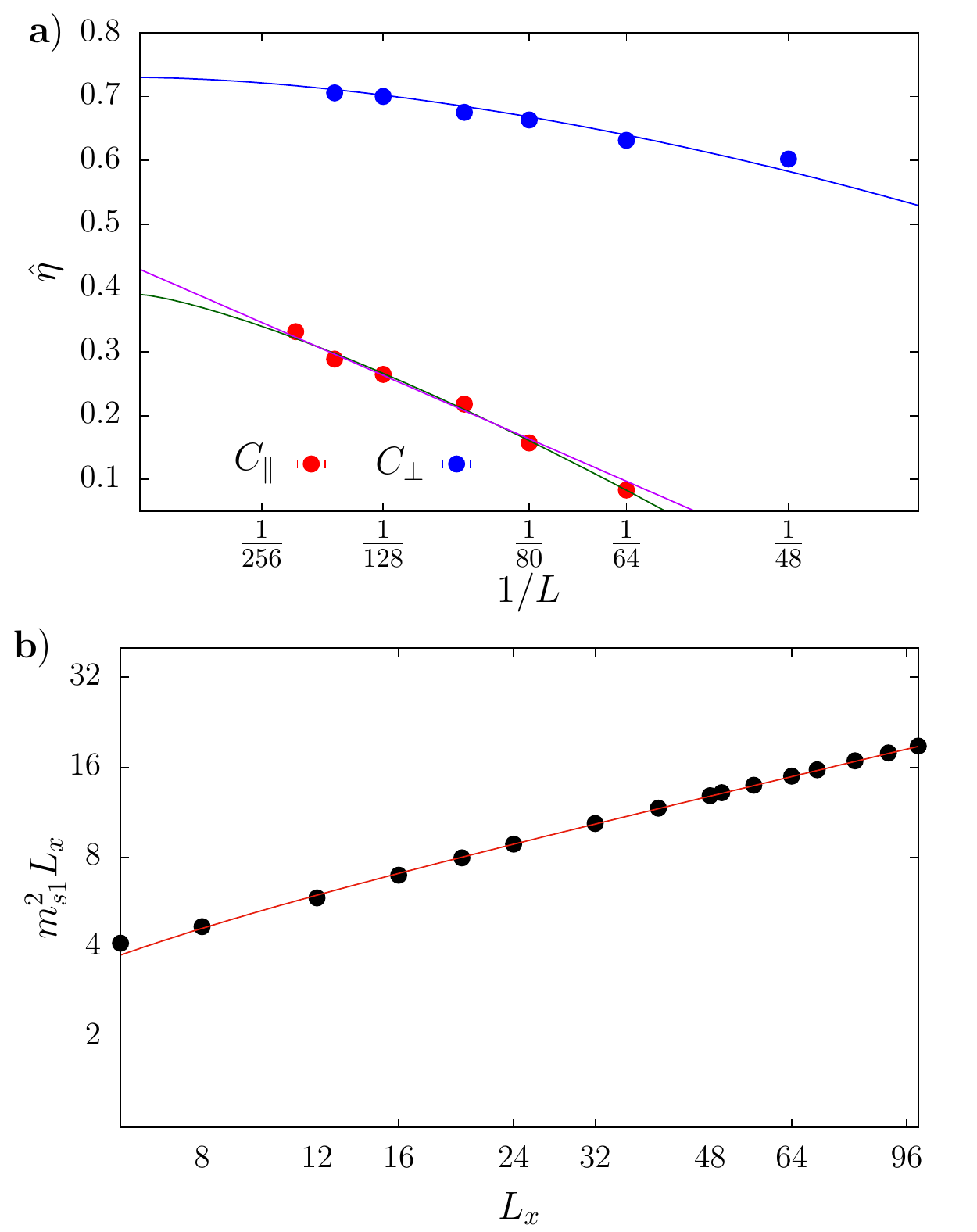}
\caption{ 
	Non-ordinary surface transition on the surfaces of the configuration Fig. \ref{fig:lat}(a). 
	(a) Finite-size dependence of the effective surface anonymous exponent $\hat{\eta}(L)$ for 
	$C_\perp$ (blue symbols) and $C_\parallel$ (red symbols). 
	The blue line shows a fit to $\hat{\eta}$ of $C_\perp$. The red and dark-green
	lines show two fits to $\hat{\eta}$ of $C_\parallel$. 
	(b) The rescaled squared magnetization of surface spins $m_{s1}^2L_x$.
	The curve is a fit to the expected scaling.
\label{fig:S1Ac2gc}}
\end{figure}

We now study the critical behavior of the surfaces along the x direction exposed by cutting a row of $J_y$ bonds, as illustrated by Fig. \ref{fig:lat}(a).
In Sec. \ref{spt} we have shown that the surface state of the x surface is gapless even in the quantum disordered SPT phase.
Together with the gapless critical mode induced by the critical bulk, a special SCB is expected.
We calculate the surface magnetization $m_{s1}^2(L_x)$, the spin correlations $C_\parallel(L_x/2)$ and $C_\perp(L_y/2)$ for systems with 
x surfaces at the bulk critical point $g_c$. 

The numerical results of $C_\parallel(L_x/2)$ and $C_\perp(L_y/2)$ as functions of system size are shown in Fig. \ref{fig:corr_xsurf}. 
Power-law decay of $C_\parallel (L_x/2)$ and $C_\perp(L_y/2)$ with system size $L_x(L_y)$ are observed.

We now apply finite-size scaling analysis to find out the surface anomalous exponents. 

We start with fitting the scaling ansatz Eq. (\ref{eqn:scal3}) to the finite-size data of $C_\perp(L_y/2)$.  
First we only use data of large enough system sizes, expecting the correction to scaling term can be discarded. 
This is true for $L\ge 200$, we find that there's no need to include the correction term in Eq. (\ref{eqn:scal3}). Our fit
finds a statistically sound estimate of $\eta_\perp = -0.27(1)$. The result is also stable upon further excluding data points of small sizes.
Still, it's tempting to include the correction to scaling term, so that we can fit more data of smaller sizes. However, it's hard to find 
the value of the effective exponent $\omega$. Setting $\omega$ to different values, e.g., $\omega=1$ or $1.5$, leads different estimates of $\eta_\perp$. 

We then define an effective exponent $\hat{\eta}$ for a pair of systems with sizes $L$ and $2L$ as in Eq. (\ref{eqn:etahat}),
with $C(L)$ stands for correlations $C_\perp$.
$\hat{\eta}(L)$ should converge to $1+\eta_\perp$ in the way of Eq. (\ref{eqn:zeff}),
according to Eq.(\ref{eqn:scal3}). 
We show the size dependence of the effective exponent $\hat{\eta}$ in Fig. \ref{fig:S1Ac2gc}(a) and fit the data to Eq. (\ref{eqn:zeff}). We find statistically
sound estimate of $\eta_\perp=-0.27(2)$ for data of $L\ge 64$, with the effective correction exponent found to be 
$\omega=1.7(5)$.
This value of $\omega$ is different from that in other models studied thus far
\cite{zhanglong,Ding2018,Weber1,Weber2}, where $\omega$ is found (or simply set) to be 1. 
The obtained results of $\eta_\perp$ are listed in Tab. \ref{tab_exp}.

The rescaled surface magnetization $m_{s1}^2 L_x$  as function of system size $L_x$ are graphed on a log-log scale in Fig. \ref{fig:S1Ac2gc}(b),
showing a power-law behavior, as expected in Eq. (\ref{eqn:scal2}) with the constant $c=0$.  
Using data of system sizes larger than $L=200$, 
we obtain statistically sound fit to Eq. (\ref{eqn:scal2}) with
the correction to scaling term discarded, since the system sizes are sufficiently large.  
The fit leads to $y_{h1}=1.761(2)$. The result is stable upon further excluding more data points. 
If we include the correction to scaling term,  we can fit more data of smaller sizes to the scaling form Eq.(\ref{eqn:scal2}).
In this case the fitting results are not sensitive to the value of $\omega$ used. Statistically consistent estimates
of $y_{h1}$ are obtained.
The best fit is $y_{h1}=1.758(2)$ with $\omega=1.5$ for data of $L \ge 128$. 
The final estimate is listed in Tab. \ref{tab_exp}.

We now turn to fit the data of $C_\parallel$ to the scaling ansatz Eq. (\ref{eqn:scal1}). First we only use data of large enough system sizes, 
expecting the correction to scaling term can be discarded.  
With data of $L\ge 256$, this leads to  $\eta_\parallel=-0.67(2)$. However, further discarding small system sizes leads to changing of the estimate of 
$\eta_\parallel$.  This suggests that the correction to scaling term can not be dropped even for sizes larger than 256.

We thus include the correction term with $\omega$ setting to 1 and 1.5, respectively, in Eq. (\ref{eqn:scal1}) and fit the data to it. We can obtain 
statistically sound fits for data of system size $L\ge 160$ for both values of $\omega$. However, the obtained $\eta_\parallel$ are different and still 
flow  with the ranges of the system sizes used in the fits.

We then define an effective exponent $\hat{\eta}$ for a pair of systems with sizes $L$ and $2L$ as in Eq. (\ref{eqn:etahat}),
with $C(L)$ stands for correlations $C_\parallel$.
According to Eq.(\ref{eqn:scal1}), $\hat{\eta}(L)$ should converge to $1+\eta_\parallel$ in the way of Eq. (\ref{eqn:zeff}).
The size dependent effective exponent $\hat{\eta}$ are graphed in Fig. \ref{fig:S1Ac2gc}(a). Fitting the data to Eq. (\ref{eqn:zeff}), we find statistically
sound estimate of $\eta_\parallel=-0.61(3)$ for data of $L\ge 64$, with the effective correction exponent $\omega=1.3(2)$. This fit is shown as the dark-green line 
in Fig. \ref{fig:S1Ac2gc}(a). 
The value of $\omega$ is slightly different from 1 that was assumed in other models studied thus far\cite{zhanglong,Ding2018,Weber1,Weber2}. 
However, if we fixed $\omega=1$ and restricted to system sizes $L\ge 100$, we find $\eta_\parallel=-0.57(2)$ with reduced $\chi^2=2.5$ and P-value $P=0.11$.  
The fit is shown as the red line in Fig. \ref{fig:S1Ac2gc}(a). 
This last estimate of $\eta_\parallel$  obey the scaling relations Eqs. (\ref{eqn:relation1}) and (\ref{eqn:relation2}) with other exponents obtained above. 
It also agrees with or is close to results found at the dangling 
surfaces of other models with dimerized Hamiltonian\cite{Ding2018, Weber1,Weber2}, where gapless surface state and the bulk critical 
state coexists, and the multicritical point is named special transition or nonordinary transition. We therefore choose this value as the 
final estimate, which is listed in Tab. \ref{tab_exp}.

It was first noticed by Ding et al \cite{Ding2018} that the exponents found in the nonordinary transitions mentioned above agree surprisingly to  
the field theoretic prediction with $\epsilon$ expansion ($\epsilon=4-d)$ to the special SCB of d-dimensional O(n) models by setting 
$\epsilon=1$ and $n=3$, \cite{Diehl, DD2} even though the 3D O(3) model does not feature such a multicritical point, as discussed in 
Sec. \ref{intro}.
However, such SCB  was claimed less universal because of slightly variation when perturbing the surface. \cite{Weber1, Weber2} 

Although a previous work attributes the phenomena to the gapless surface state of the SPT bulk state \cite{zhanglong},
the role of SPT was challenged by the fact that the non-dangling edge shows ordinary SCB at the same AKLT to N\'eel QCP. 
It was further opposed by the finding that the suggested AKLT state is not a real SPT phase. \cite{Weber1}
Our model shows a genuine SPT phase separated from the N\'eel phase by a QCP in the 3D O(3) universality class.
Our results provide evidences of nonordinary SCB induced by SPT physics in the surface perpendicular to the spin-1 chains. 
In particular, our model does not have a dangling edge or surface, instead, 
the surface we study here is not dangling at all, still we find the
nonordinary SCBs at the bulk 3D O(3) critical point. We therefore believe the SPT physics is essential here.

\section{Conclusions}
\label{concl}

The discovery of nonordinary surface transitions at bulk quantum critical point, that is precluded in corresponding 
classical critical point, has inspired interests in the investigations on the quantum origin of such behaviors.
So far researches were focused on dimerized antiferromagnetic Heisenberg models and it was noticed that surfaces formed by 
dangling spins in the dimerized models show such nonordinary SCBs. However, the role of SPT and spin-1/2 related 
topological terms  were still under debate.

We have studied the system of coupled Haldane chains in two-dimensional 
square lattice with interchain couplings, which has an SPT Haldane phase when the interchain couplings are weak. 
Increasing interchain couplings, the model enters a N\'eel phase through a quantum critical point belonging to the 3D
O(3) universality class. 
Different from previously studied dimerized models, this model does not have surfaces formed by dangling or non-dangling 
spins, therefore offers an opportunity to test the role of SPT in the origin of nonordinary SCBs.
In particular, the model is anisotropic. The surfaces along the spin-1 chain direction (x surfaces) are completely different from the 
surfaces perpendicular (y surfaces) to the chain direction.

Using unbiased quantum Monte Carlo simulations, we have studied in a great detail the string order parameter 
of the gapped Haldane phase and its surface states. We found that, although the string order decay exponentially with system 
size, as interchain interactions are introduced, the x surface states are gapless even in the gapped Haldane phase while the y 
surface is gapped. This verified that the gapless surface state is the property of a SPT phase.

We have then studied the surface critical behaviors of surfaces along the x and the y direction at the bulk critical point, by 
calculating the spin-spin correlations along the surface and/or perpendicular to the surface and the surface magnetization. 
Ordinary SCBs were found at the surfaces along the spin-1 chain direction, which have gapped surface states in the 
gapped SPT phase. We found that gapless surface states at the surfaces perpendicular to the spin-1 chains in the 
gapped SPT phase result in multicritical special transitions at the quantum critical point, with exponents in agreement with 
those found in other SCBs at dangling surfaces of dimerized models. 
This behavior is of pure quantum origin, because nonordinary SCB is precluded in the 3D classical O(3) models.
All exponents found in the surface transitions, either ordinary or special,  satisfy the scaling relations.

Although the quantum origin of the nonordinary SCB in dimerized Heisenberg models seems not induced by the SPT physics in 
particular, and the mechanism of the nonordinary SCBs in these models is still not clear,
our numerical results support that the SPT and its gapless surface states together with the gapless critical modes 
at least offers one quantum origin of nonordinary SCB.

\begin{acknowledgments}
W.Z and W.G were supported by the National Natural Science Foundation of China under Grant No.~11775021 and No.~11734002. C.D. was 
supported by the National Natural Science Foundation of China under Grant No.~11975024 and the Anhui Provincial Supporting Program 
for Excellent Young Talents in Colleges and Universities under Grant No.~gxyqZD2019023. L.Z. was supported by the National Natural 
Science Foundation of China under Grant No.~11804337 and the CAS Youth Innovation Promotion Association. The authors acknowledge 
supporting extended by the Super Computing Center of Beijing Normal University. 
\end{acknowledgments}

\bibliography{my}

\begin{thebibliography}{54}%
\makeatletter
\providecommand \@ifxundefined [1]{%
 \@ifx{#1\undefined}
}%
\providecommand \@ifnum [1]{%
 \ifnum #1\expandafter \@firstoftwo
 \else \expandafter \@secondoftwo
 \fi
}%
\providecommand \@ifx [1]{%
 \ifx #1\expandafter \@firstoftwo
 \else \expandafter \@secondoftwo
 \fi
}%
\providecommand \natexlab [1]{#1}%
\providecommand \enquote  [1]{``#1''}%
\providecommand \bibnamefont  [1]{#1}%
\providecommand \bibfnamefont [1]{#1}%
\providecommand \citenamefont [1]{#1}%
\providecommand \href@noop [0]{\@secondoftwo}%
\providecommand \href [0]{\begingroup \@sanitize@url \@href}%
\providecommand \@href[1]{\@@startlink{#1}\@@href}%
\providecommand \@@href[1]{\endgroup#1\@@endlink}%
\providecommand \@sanitize@url [0]{\catcode `\\12\catcode `\$12\catcode
  `\&12\catcode `\#12\catcode `\^12\catcode `\_12\catcode `\%12\relax}%
\providecommand \@@startlink[1]{}%
\providecommand \@@endlink[0]{}%
\providecommand \url  [0]{\begingroup\@sanitize@url \@url }%
\providecommand \@url [1]{\endgroup\@href {#1}{\urlprefix }}%
\providecommand \urlprefix  [0]{URL }%
\providecommand \Eprint [0]{\href }%
\providecommand \doibase [0]{https://doi.org/}%
\providecommand \selectlanguage [0]{\@gobble}%
\providecommand \bibinfo  [0]{\@secondoftwo}%
\providecommand \bibfield  [0]{\@secondoftwo}%
\providecommand \translation [1]{[#1]}%
\providecommand \BibitemOpen [0]{}%
\providecommand \bibitemStop [0]{}%
\providecommand \bibitemNoStop [0]{.\EOS\space}%
\providecommand \EOS [0]{\spacefactor3000\relax}%
\providecommand \BibitemShut  [1]{\csname bibitem#1\endcsname}%
\let\auto@bib@innerbib\@empty
\bibitem [{\citenamefont {Binder}(1983)}]{binder}%
  \BibitemOpen
  \bibfield  {author} {\bibinfo {author} {\bibfnamefont {K.}~\bibnamefont
  {Binder}},\ }\href
  {https://www.amazon.com/Phase-Transitions-Critical-Phenomena-8/dp/0122203089}
  {\emph {\bibinfo {title} {Phase Transitions and Critical Phenomena}}},\
  edited by\ \bibinfo {editor} {\bibfnamefont {C.}~\bibnamefont {Domb}}\ and\
  \bibinfo {editor} {\bibfnamefont {J.~L.}\ \bibnamefont {Lebowitz}},\
  Vol.~\bibinfo {volume} {8}\ (\bibinfo  {publisher} {Academic Press},\
  \bibinfo {address} {London},\ \bibinfo {year} {1983})\BibitemShut {NoStop}%
\bibitem [{\citenamefont {Binder}\ and\ \citenamefont {Hohenberg}(1974)}]{BH}%
  \BibitemOpen
  \bibfield  {author} {\bibinfo {author} {\bibfnamefont {K.}~\bibnamefont
  {Binder}}\ and\ \bibinfo {author} {\bibfnamefont {P.~C.}\ \bibnamefont
  {Hohenberg}},\ }\bibfield  {title} {\bibinfo {title} {Surface effects on
  magnetic phase transitions},\ }\href
  {https://doi.org/10.1103/PhysRevB.9.2194} {\bibfield  {journal} {\bibinfo
  {journal} {Phys. Rev. B}\ }\textbf {\bibinfo {volume} {9}},\ \bibinfo {pages}
  {2194} (\bibinfo {year} {1974})}\BibitemShut {NoStop}%
\bibitem [{\citenamefont {Binder}(1986)}]{Diehl}%
  \BibitemOpen
  \bibfield  {author} {\bibinfo {author} {\bibfnamefont {K.}~\bibnamefont
  {Binder}},\ }\href
  {https://www.amazon.com/Phase-Transitions-Critical-Phenomena-8/dp/0122203089}
  {\emph {\bibinfo {title} {Phase Transitions and Critical Phenomena}}},\
  edited by\ \bibinfo {editor} {\bibfnamefont {C.}~\bibnamefont {Domb}}\ and\
  \bibinfo {editor} {\bibfnamefont {J.~L.}\ \bibnamefont {Lebowitz}},\
  Vol.~\bibinfo {volume} {10}\ (\bibinfo  {publisher} {Academic Press},\
  \bibinfo {address} {London},\ \bibinfo {year} {1986})\BibitemShut {NoStop}%
\bibitem [{\citenamefont {Mermin}\ and\ \citenamefont {Wagner}(1966)}]{Mermin}%
  \BibitemOpen
  \bibfield  {author} {\bibinfo {author} {\bibfnamefont {N.~D.}\ \bibnamefont
  {Mermin}}\ and\ \bibinfo {author} {\bibfnamefont {H.}~\bibnamefont
  {Wagner}},\ }\bibfield  {title} {\bibinfo {title} {Absence of ferromagnetism
  or antiferromagnetism in one- or two-dimensional isotropic heisenberg
  models},\ }\href {https://doi.org/10.1103/PhysRevLett.17.1133} {\bibfield
  {journal} {\bibinfo  {journal} {Phys. Rev. Lett.}\ }\textbf {\bibinfo
  {volume} {17}},\ \bibinfo {pages} {1133} (\bibinfo {year}
  {1966})}\BibitemShut {NoStop}%
\bibitem [{\citenamefont {Sachdev}(2011)}]{Sachdev}%
  \BibitemOpen
  \bibfield  {author} {\bibinfo {author} {\bibfnamefont {S.}~\bibnamefont
  {Sachdev}},\ }\href {https://doi.org/10.1017/CBO9780511973765} {\emph
  {\bibinfo {title} {Quantum phase transitions}}}\ (\bibinfo  {publisher}
  {Cambridge University Press, Cambridge, UK},\ \bibinfo {year}
  {2011})\BibitemShut {NoStop}%
\bibitem [{\citenamefont {Affleck}\ \emph {et~al.}(1987)\citenamefont
  {Affleck}, \citenamefont {Kennedy}, \citenamefont {Lieb},\ and\ \citenamefont
  {Tasaki}}]{AKLT}%
  \BibitemOpen
  \bibfield  {author} {\bibinfo {author} {\bibfnamefont {I.}~\bibnamefont
  {Affleck}}, \bibinfo {author} {\bibfnamefont {T.}~\bibnamefont {Kennedy}},
  \bibinfo {author} {\bibfnamefont {E.~H.}\ \bibnamefont {Lieb}},\ and\
  \bibinfo {author} {\bibfnamefont {H.}~\bibnamefont {Tasaki}},\ }\bibfield
  {title} {\bibinfo {title} {Rigorous results on valence-bond ground states in
  antiferromagnets},\ }\href {https://doi.org/10.1103/PhysRevLett.59.799}
  {\bibfield  {journal} {\bibinfo  {journal} {Phys. Rev. Lett.}\ }\textbf
  {\bibinfo {volume} {59}},\ \bibinfo {pages} {799} (\bibinfo {year}
  {1987})}\BibitemShut {NoStop}%
\bibitem [{\citenamefont {Zhang}\ and\ \citenamefont {Wang}(2017)}]{zhanglong}%
  \BibitemOpen
  \bibfield  {author} {\bibinfo {author} {\bibfnamefont {L.}~\bibnamefont
  {Zhang}}\ and\ \bibinfo {author} {\bibfnamefont {F.}~\bibnamefont {Wang}},\
  }\bibfield  {title} {\bibinfo {title} {Unconventional surface critical
  behavior induced by a quantum phase transition from the two-dimensional
  affleck-kennedy-lieb-tasaki phase to a n\'eel-ordered phase},\ }\href
  {https://doi.org/10.1103/PhysRevLett.118.087201} {\bibfield  {journal}
  {\bibinfo  {journal} {Phys. Rev. Lett.}\ }\textbf {\bibinfo {volume} {118}},\
  \bibinfo {pages} {087201} (\bibinfo {year} {2017})}\BibitemShut {NoStop}%
\bibitem [{\citenamefont {Gu}\ and\ \citenamefont {Wen}(2009)}]{GuWen}%
  \BibitemOpen
  \bibfield  {author} {\bibinfo {author} {\bibfnamefont {Z.-C.}\ \bibnamefont
  {Gu}}\ and\ \bibinfo {author} {\bibfnamefont {X.-G.}\ \bibnamefont {Wen}},\
  }\bibfield  {title} {\bibinfo {title} {Tensor-entanglement-filtering
  renormalization approach and symmetry-protected topological order},\ }\href
  {https://doi.org/10.1103/PhysRevB.80.155131} {\bibfield  {journal} {\bibinfo
  {journal} {Phys. Rev. B}\ }\textbf {\bibinfo {volume} {80}},\ \bibinfo
  {pages} {155131} (\bibinfo {year} {2009})}\BibitemShut {NoStop}%
\bibitem [{\citenamefont {Pollmann}\ \emph {et~al.}(2010)\citenamefont
  {Pollmann}, \citenamefont {Turner}, \citenamefont {Berg},\ and\ \citenamefont
  {Oshikawa}}]{Pollmann}%
  \BibitemOpen
  \bibfield  {author} {\bibinfo {author} {\bibfnamefont {F.}~\bibnamefont
  {Pollmann}}, \bibinfo {author} {\bibfnamefont {A.~M.}\ \bibnamefont
  {Turner}}, \bibinfo {author} {\bibfnamefont {E.}~\bibnamefont {Berg}},\ and\
  \bibinfo {author} {\bibfnamefont {M.}~\bibnamefont {Oshikawa}},\ }\bibfield
  {title} {\bibinfo {title} {Entanglement spectrum of a topological phase in
  one dimension},\ }\href {https://doi.org/10.1103/PhysRevB.81.064439}
  {\bibfield  {journal} {\bibinfo  {journal} {Phys. Rev. B}\ }\textbf {\bibinfo
  {volume} {81}},\ \bibinfo {pages} {064439} (\bibinfo {year}
  {2010})}\BibitemShut {NoStop}%
\bibitem [{\citenamefont {Ding}\ \emph {et~al.}(2018)\citenamefont {Ding},
  \citenamefont {Zhang},\ and\ \citenamefont {Guo}}]{Ding2018}%
  \BibitemOpen
  \bibfield  {author} {\bibinfo {author} {\bibfnamefont {C.}~\bibnamefont
  {Ding}}, \bibinfo {author} {\bibfnamefont {L.}~\bibnamefont {Zhang}},\ and\
  \bibinfo {author} {\bibfnamefont {W.}~\bibnamefont {Guo}},\ }\bibfield
  {title} {\bibinfo {title} {Engineering surface critical behavior of
  ($2+1$)-dimensional o(3) quantum critical points},\ }\href
  {https://doi.org/10.1103/PhysRevLett.120.235701} {\bibfield  {journal}
  {\bibinfo  {journal} {Phys. Rev. Lett.}\ }\textbf {\bibinfo {volume} {120}},\
  \bibinfo {pages} {235701} (\bibinfo {year} {2018})}\BibitemShut {NoStop}%
\bibitem [{\citenamefont {Weber}\ \emph {et~al.}(2018)\citenamefont {Weber},
  \citenamefont {Parisen~Toldin},\ and\ \citenamefont {Wessel}}]{Weber1}%
  \BibitemOpen
  \bibfield  {author} {\bibinfo {author} {\bibfnamefont {L.}~\bibnamefont
  {Weber}}, \bibinfo {author} {\bibfnamefont {F.}~\bibnamefont
  {Parisen~Toldin}},\ and\ \bibinfo {author} {\bibfnamefont {S.}~\bibnamefont
  {Wessel}},\ }\bibfield  {title} {\bibinfo {title} {Nonordinary edge
  criticality of two-dimensional quantum critical magnets},\ }\href
  {https://doi.org/10.1103/PhysRevB.98.140403} {\bibfield  {journal} {\bibinfo
  {journal} {Phys. Rev. B}\ }\textbf {\bibinfo {volume} {98}},\ \bibinfo
  {pages} {140403} (\bibinfo {year} {2018})}\BibitemShut {NoStop}%
\bibitem [{\citenamefont {Weber}\ and\ \citenamefont {Wessel}(2019)}]{Weber2}%
  \BibitemOpen
  \bibfield  {author} {\bibinfo {author} {\bibfnamefont {L.}~\bibnamefont
  {Weber}}\ and\ \bibinfo {author} {\bibfnamefont {S.}~\bibnamefont {Wessel}},\
  }\bibfield  {title} {\bibinfo {title} {Nonordinary criticality at the edges
  of planar spin-1 heisenberg antiferromagnets},\ }\href
  {https://doi.org/10.1103/PhysRevB.100.054437} {\bibfield  {journal} {\bibinfo
   {journal} {Phys. Rev. B}\ }\textbf {\bibinfo {volume} {100}},\ \bibinfo
  {pages} {054437} (\bibinfo {year} {2019})}\BibitemShut {NoStop}%
\bibitem [{\citenamefont {Mutka}\ \emph {et~al.}(1991)\citenamefont {Mutka},
  \citenamefont {Payen}, \citenamefont {Molini\'e}, \citenamefont {Soubeyroux},
  \citenamefont {Colombet},\ and\ \citenamefont {Taylor}}]{Mutka}%
  \BibitemOpen
  \bibfield  {author} {\bibinfo {author} {\bibfnamefont {H.}~\bibnamefont
  {Mutka}}, \bibinfo {author} {\bibfnamefont {C.}~\bibnamefont {Payen}},
  \bibinfo {author} {\bibfnamefont {P.}~\bibnamefont {Molini\'e}}, \bibinfo
  {author} {\bibfnamefont {J.~L.}\ \bibnamefont {Soubeyroux}}, \bibinfo
  {author} {\bibfnamefont {P.}~\bibnamefont {Colombet}},\ and\ \bibinfo
  {author} {\bibfnamefont {A.~D.}\ \bibnamefont {Taylor}},\ }\bibfield  {title}
  {\bibinfo {title} {Dynamic structure factor [s(q,\ensuremath{\omega})] of the
  s=1 quasi-one-dimensional heisenberg antiferromagnet: Neutron-scattering
  study on ${\mathrm{agvp}}_{2}$${\mathrm{s}}_{6}$},\ }\href
  {https://doi.org/10.1103/PhysRevLett.67.497} {\bibfield  {journal} {\bibinfo
  {journal} {Phys. Rev. Lett.}\ }\textbf {\bibinfo {volume} {67}},\ \bibinfo
  {pages} {497} (\bibinfo {year} {1991})}\BibitemShut {NoStop}%
\bibitem [{\citenamefont {Asano}\ \emph {et~al.}(1994)\citenamefont {Asano},
  \citenamefont {Ajiro}, \citenamefont {Mekata}, \citenamefont {Yamazaki},
  \citenamefont {Hosoito}, \citenamefont {Shinjo},\ and\ \citenamefont
  {Kikuchi}}]{Asano}%
  \BibitemOpen
  \bibfield  {author} {\bibinfo {author} {\bibfnamefont {T.}~\bibnamefont
  {Asano}}, \bibinfo {author} {\bibfnamefont {Y.}~\bibnamefont {Ajiro}},
  \bibinfo {author} {\bibfnamefont {M.}~\bibnamefont {Mekata}}, \bibinfo
  {author} {\bibfnamefont {H.}~\bibnamefont {Yamazaki}}, \bibinfo {author}
  {\bibfnamefont {N.}~\bibnamefont {Hosoito}}, \bibinfo {author} {\bibfnamefont
  {T.}~\bibnamefont {Shinjo}},\ and\ \bibinfo {author} {\bibfnamefont
  {H.}~\bibnamefont {Kikuchi}},\ }\bibfield  {title} {\bibinfo {title} {Single
  crystal susceptibility of the s = 1 one-dimensional heisenberg
  antiferromagnet agvp2s6},\ }\href
  {https://doi.org/https://doi.org/10.1016/0038-1098(94)90944-X} {\bibfield
  {journal} {\bibinfo  {journal} {Solid State Communications}\ }\textbf
  {\bibinfo {volume} {90}},\ \bibinfo {pages} {125 } (\bibinfo {year}
  {1994})}\BibitemShut {NoStop}%
\bibitem [{\citenamefont {Honda}\ \emph {et~al.}(2001)\citenamefont {Honda},
  \citenamefont {Katsumata}, \citenamefont {Nishiyama},\ and\ \citenamefont
  {Harada}}]{Honda}%
  \BibitemOpen
  \bibfield  {author} {\bibinfo {author} {\bibfnamefont {Z.}~\bibnamefont
  {Honda}}, \bibinfo {author} {\bibfnamefont {K.}~\bibnamefont {Katsumata}},
  \bibinfo {author} {\bibfnamefont {Y.}~\bibnamefont {Nishiyama}},\ and\
  \bibinfo {author} {\bibfnamefont {I.}~\bibnamefont {Harada}},\ }\bibfield
  {title} {\bibinfo {title} {Field-induced long-range ordering in an $s=1$
  quasi-one-dimensional heisenberg antiferromagnet},\ }\href
  {https://doi.org/10.1103/PhysRevB.63.064420} {\bibfield  {journal} {\bibinfo
  {journal} {Phys. Rev. B}\ }\textbf {\bibinfo {volume} {63}},\ \bibinfo
  {pages} {064420} (\bibinfo {year} {2001})}\BibitemShut {NoStop}%
\bibitem [{\citenamefont {Renard}\ \emph {et~al.}(1987)\citenamefont {Renard},
  \citenamefont {Verdaguer}, \citenamefont {Regnault}, \citenamefont
  {Erkelens}, \citenamefont {Rossat-Mignod},\ and\ \citenamefont
  {Stirling}}]{Renard1}%
  \BibitemOpen
  \bibfield  {author} {\bibinfo {author} {\bibfnamefont {J.~P.}\ \bibnamefont
  {Renard}}, \bibinfo {author} {\bibfnamefont {M.}~\bibnamefont {Verdaguer}},
  \bibinfo {author} {\bibfnamefont {L.~P.}\ \bibnamefont {Regnault}}, \bibinfo
  {author} {\bibfnamefont {W.~A.~C.}\ \bibnamefont {Erkelens}}, \bibinfo
  {author} {\bibfnamefont {J.}~\bibnamefont {Rossat-Mignod}},\ and\ \bibinfo
  {author} {\bibfnamefont {W.~G.}\ \bibnamefont {Stirling}},\ }\bibfield
  {title} {\bibinfo {title} {Presumption for a quantum energy gap in the
  quasi-one-dimensional s = 1 heisenberg antiferromagnet
  ni(c2h8n2)2no2(clo4)},\ }\href {https://doi.org/10.1209/0295-5075/3/8/013}
  {\bibfield  {journal} {\bibinfo  {journal} {Europhysics Letters ({EPL})}\
  }\textbf {\bibinfo {volume} {3}},\ \bibinfo {pages} {945} (\bibinfo {year}
  {1987})}\BibitemShut {NoStop}%
\bibitem [{\citenamefont {Renard}\ \emph {et~al.}(1988)\citenamefont {Renard},
  \citenamefont {Verdaguer}, \citenamefont {Regnault}, \citenamefont
  {Erkelens}, \citenamefont {Rossat‐Mignod}, \citenamefont {Ribas},
  \citenamefont {Stirling},\ and\ \citenamefont {Vettier}}]{Renard2}%
  \BibitemOpen
  \bibfield  {author} {\bibinfo {author} {\bibfnamefont {J.~P.}\ \bibnamefont
  {Renard}}, \bibinfo {author} {\bibfnamefont {M.}~\bibnamefont {Verdaguer}},
  \bibinfo {author} {\bibfnamefont {L.~P.}\ \bibnamefont {Regnault}}, \bibinfo
  {author} {\bibfnamefont {W.~A.~C.}\ \bibnamefont {Erkelens}}, \bibinfo
  {author} {\bibfnamefont {J.}~\bibnamefont {Rossat‐Mignod}}, \bibinfo
  {author} {\bibfnamefont {J.}~\bibnamefont {Ribas}}, \bibinfo {author}
  {\bibfnamefont {W.~G.}\ \bibnamefont {Stirling}},\ and\ \bibinfo {author}
  {\bibfnamefont {C.}~\bibnamefont {Vettier}},\ }\bibfield  {title} {\bibinfo
  {title} {Quantum energy gap in two quasi‐one‐dimensional s=1 heisenberg
  antiferromagnets (invited)},\ }\href {https://doi.org/10.1063/1.340736}
  {\bibfield  {journal} {\bibinfo  {journal} {Journal of Applied Physics}\
  }\textbf {\bibinfo {volume} {63}},\ \bibinfo {pages} {3538} (\bibinfo {year}
  {1988})}\BibitemShut {NoStop}%
\bibitem [{\citenamefont {Regnault}\ \emph {et~al.}(1994)\citenamefont
  {Regnault}, \citenamefont {Zaliznyak}, \citenamefont {Renard},\ and\
  \citenamefont {Vettier}}]{Regnault}%
  \BibitemOpen
  \bibfield  {author} {\bibinfo {author} {\bibfnamefont {L.~P.}\ \bibnamefont
  {Regnault}}, \bibinfo {author} {\bibfnamefont {I.}~\bibnamefont {Zaliznyak}},
  \bibinfo {author} {\bibfnamefont {J.~P.}\ \bibnamefont {Renard}},\ and\
  \bibinfo {author} {\bibfnamefont {C.}~\bibnamefont {Vettier}},\ }\bibfield
  {title} {\bibinfo {title} {Inelastic-neutron-scattering study of the spin
  dynamics in the haldane-gap system
  ni(${\mathrm{c}}_{2}$${\mathrm{h}}_{8}$${\mathrm{n}}_{2}$${)}_{2}$${\mathrm{no}}_{2}$${\rm{clo}}_{4}$},\
  }\href {https://doi.org/10.1103/PhysRevB.50.9174} {\bibfield  {journal}
  {\bibinfo  {journal} {Phys. Rev. B}\ }\textbf {\bibinfo {volume} {50}},\
  \bibinfo {pages} {9174} (\bibinfo {year} {1994})}\BibitemShut {NoStop}%
\bibitem [{\citenamefont {Takigawa}\ \emph {et~al.}(1995)\citenamefont
  {Takigawa}, \citenamefont {Asano}, \citenamefont {Ajiro},\ and\ \citenamefont
  {Mekata}}]{Takigawa1}%
  \BibitemOpen
  \bibfield  {author} {\bibinfo {author} {\bibfnamefont {M.}~\bibnamefont
  {Takigawa}}, \bibinfo {author} {\bibfnamefont {T.}~\bibnamefont {Asano}},
  \bibinfo {author} {\bibfnamefont {Y.}~\bibnamefont {Ajiro}},\ and\ \bibinfo
  {author} {\bibfnamefont {M.}~\bibnamefont {Mekata}},\ }\bibfield  {title}
  {\bibinfo {title} {Static properties of the s=1 one-dimensional
  antiferromagnet ${\mathrm{agvp}}_{2}$${\mathrm{s}}_{6}$},\ }\href
  {https://doi.org/10.1103/PhysRevB.52.R13087} {\bibfield  {journal} {\bibinfo
  {journal} {Phys. Rev. B}\ }\textbf {\bibinfo {volume} {52}},\ \bibinfo
  {pages} {R13087} (\bibinfo {year} {1995})}\BibitemShut {NoStop}%
\bibitem [{\citenamefont {Takigawa}\ \emph {et~al.}(1996)\citenamefont
  {Takigawa}, \citenamefont {Asano}, \citenamefont {Ajiro}, \citenamefont
  {Mekata},\ and\ \citenamefont {Uemura}}]{Takigawa2}%
  \BibitemOpen
  \bibfield  {author} {\bibinfo {author} {\bibfnamefont {M.}~\bibnamefont
  {Takigawa}}, \bibinfo {author} {\bibfnamefont {T.}~\bibnamefont {Asano}},
  \bibinfo {author} {\bibfnamefont {Y.}~\bibnamefont {Ajiro}}, \bibinfo
  {author} {\bibfnamefont {M.}~\bibnamefont {Mekata}},\ and\ \bibinfo {author}
  {\bibfnamefont {Y.~J.}\ \bibnamefont {Uemura}},\ }\bibfield  {title}
  {\bibinfo {title} {Dynamics in the
  $\mathit{S}\phantom{\rule{0ex}{0ex}}=\phantom{\rule{0ex}{0ex}}1$
  one-dimensional antiferromagnet agv${\mathrm{p}}_{2}$${\mathrm{s}}_{6}$ via
  ${}^{31}$p and ${}^{51}$v nmr},\ }\href
  {https://doi.org/10.1103/PhysRevLett.76.2173} {\bibfield  {journal} {\bibinfo
   {journal} {Phys. Rev. Lett.}\ }\textbf {\bibinfo {volume} {76}},\ \bibinfo
  {pages} {2173} (\bibinfo {year} {1996})}\BibitemShut {NoStop}%
\bibitem [{\citenamefont {Zaliznyak}\ \emph {et~al.}(1998)\citenamefont
  {Zaliznyak}, \citenamefont {Dender}, \citenamefont {Broholm},\ and\
  \citenamefont {Reich}}]{Zaliznyak}%
  \BibitemOpen
  \bibfield  {author} {\bibinfo {author} {\bibfnamefont {I.~A.}\ \bibnamefont
  {Zaliznyak}}, \bibinfo {author} {\bibfnamefont {D.~C.}\ \bibnamefont
  {Dender}}, \bibinfo {author} {\bibfnamefont {C.}~\bibnamefont {Broholm}},\
  and\ \bibinfo {author} {\bibfnamefont {D.~H.}\ \bibnamefont {Reich}},\
  }\bibfield  {title} {\bibinfo {title} {Tuning the spin hamiltonian of
  $\mathrm{Ni}({\mathrm{c}}_{2}{\mathrm{h}}_{8}{\mathrm{n}}_{2}{)}_{2}{\mathrm{no}}_{2}{\mathrm{clo}}_{4}$
  by external pressure: A neutron-scattering study},\ }\href
  {https://doi.org/10.1103/PhysRevB.57.5200} {\bibfield  {journal} {\bibinfo
  {journal} {Phys. Rev. B}\ }\textbf {\bibinfo {volume} {57}},\ \bibinfo
  {pages} {5200} (\bibinfo {year} {1998})}\BibitemShut {NoStop}%
\bibitem [{\citenamefont {Uchiyama}\ \emph {et~al.}(1999)\citenamefont
  {Uchiyama}, \citenamefont {Sasago}, \citenamefont {Tsukada}, \citenamefont
  {Uchinokura}, \citenamefont {Zheludev}, \citenamefont {Hayashi},
  \citenamefont {Miura},\ and\ \citenamefont {B\"oni}}]{Uchiyama}%
  \BibitemOpen
  \bibfield  {author} {\bibinfo {author} {\bibfnamefont {Y.}~\bibnamefont
  {Uchiyama}}, \bibinfo {author} {\bibfnamefont {Y.}~\bibnamefont {Sasago}},
  \bibinfo {author} {\bibfnamefont {I.}~\bibnamefont {Tsukada}}, \bibinfo
  {author} {\bibfnamefont {K.}~\bibnamefont {Uchinokura}}, \bibinfo {author}
  {\bibfnamefont {A.}~\bibnamefont {Zheludev}}, \bibinfo {author}
  {\bibfnamefont {T.}~\bibnamefont {Hayashi}}, \bibinfo {author} {\bibfnamefont
  {N.}~\bibnamefont {Miura}},\ and\ \bibinfo {author} {\bibfnamefont
  {P.}~\bibnamefont {B\"oni}},\ }\bibfield  {title} {\bibinfo {title}
  {Spin-vacancy-induced long-range order in a new haldane-gap
  antiferromagnet},\ }\href {https://doi.org/10.1103/PhysRevLett.83.632}
  {\bibfield  {journal} {\bibinfo  {journal} {Phys. Rev. Lett.}\ }\textbf
  {\bibinfo {volume} {83}},\ \bibinfo {pages} {632} (\bibinfo {year}
  {1999})}\BibitemShut {NoStop}%
\bibitem [{\citenamefont {Pahari}\ \emph {et~al.}(2006)\citenamefont {Pahari},
  \citenamefont {Ghoshray}, \citenamefont {Sarkar}, \citenamefont
  {Bandyopadhyay},\ and\ \citenamefont {Ghoshray}}]{Pahari}%
  \BibitemOpen
  \bibfield  {author} {\bibinfo {author} {\bibfnamefont {B.}~\bibnamefont
  {Pahari}}, \bibinfo {author} {\bibfnamefont {K.}~\bibnamefont {Ghoshray}},
  \bibinfo {author} {\bibfnamefont {R.}~\bibnamefont {Sarkar}}, \bibinfo
  {author} {\bibfnamefont {B.}~\bibnamefont {Bandyopadhyay}},\ and\ \bibinfo
  {author} {\bibfnamefont {A.}~\bibnamefont {Ghoshray}},\ }\bibfield  {title}
  {\bibinfo {title} {Nmr study of $^{51}\mathrm{V}$ in quasi-one-dimensional
  integer spin chain compound
  $\mathrm{Sr}{\mathrm{ni}}_{2}{\mathrm{v}}_{2}{\mathrm{o}}_{8}$},\ }\href
  {https://doi.org/10.1103/PhysRevB.73.012407} {\bibfield  {journal} {\bibinfo
  {journal} {Phys. Rev. B}\ }\textbf {\bibinfo {volume} {73}},\ \bibinfo
  {pages} {012407} (\bibinfo {year} {2006})}\BibitemShut {NoStop}%
\bibitem [{\citenamefont {Gadet}\ \emph {et~al.}(1991)\citenamefont {Gadet},
  \citenamefont {Verdaguer}, \citenamefont {Briois}, \citenamefont {Gleizes},
  \citenamefont {Renard}, \citenamefont {Beauvillain}, \citenamefont
  {Chappert}, \citenamefont {Goto}, \citenamefont {Le~Dang},\ and\
  \citenamefont {Veillet}}]{Gadet}%
  \BibitemOpen
  \bibfield  {author} {\bibinfo {author} {\bibfnamefont {V.}~\bibnamefont
  {Gadet}}, \bibinfo {author} {\bibfnamefont {M.}~\bibnamefont {Verdaguer}},
  \bibinfo {author} {\bibfnamefont {V.}~\bibnamefont {Briois}}, \bibinfo
  {author} {\bibfnamefont {A.}~\bibnamefont {Gleizes}}, \bibinfo {author}
  {\bibfnamefont {J.~P.}\ \bibnamefont {Renard}}, \bibinfo {author}
  {\bibfnamefont {P.}~\bibnamefont {Beauvillain}}, \bibinfo {author}
  {\bibfnamefont {C.}~\bibnamefont {Chappert}}, \bibinfo {author}
  {\bibfnamefont {T.}~\bibnamefont {Goto}}, \bibinfo {author} {\bibfnamefont
  {K.}~\bibnamefont {Le~Dang}},\ and\ \bibinfo {author} {\bibfnamefont
  {P.}~\bibnamefont {Veillet}},\ }\bibfield  {title} {\bibinfo {title}
  {Structural and magnetic properties of
  (${\mathrm{ch}}_{3}$${)}_{4}$nni(${\mathrm{no}}_{2}$${)}_{3}$: A haldane-gap
  system},\ }\href {https://doi.org/10.1103/PhysRevB.44.705} {\bibfield
  {journal} {\bibinfo  {journal} {Phys. Rev. B}\ }\textbf {\bibinfo {volume}
  {44}},\ \bibinfo {pages} {705} (\bibinfo {year} {1991})}\BibitemShut
  {NoStop}%
\bibitem [{\citenamefont {Darriet}\ and\ \citenamefont
  {Regnault}(1993)}]{Darriet}%
  \BibitemOpen
  \bibfield  {author} {\bibinfo {author} {\bibfnamefont {J.}~\bibnamefont
  {Darriet}}\ and\ \bibinfo {author} {\bibfnamefont {L.}~\bibnamefont
  {Regnault}},\ }\bibfield  {title} {\bibinfo {title} {The compound y2banio5: A
  new example of a haldane gap in a s = 1 magnetic chain},\ }\href
  {https://doi.org/https://doi.org/10.1016/0038-1098(93)90455-V} {\bibfield
  {journal} {\bibinfo  {journal} {Solid State Communications}\ }\textbf
  {\bibinfo {volume} {86}},\ \bibinfo {pages} {409 } (\bibinfo {year}
  {1993})}\BibitemShut {NoStop}%
\bibitem [{\citenamefont {Haldane}(1983)}]{Haldane}%
  \BibitemOpen
  \bibfield  {author} {\bibinfo {author} {\bibfnamefont {F.~D.~M.}\
  \bibnamefont {Haldane}},\ }\bibfield  {title} {\bibinfo {title} {Nonlinear
  field theory of large-spin heisenberg antiferromagnets: Semiclassically
  quantized solitons of the one-dimensional easy-axis n\'eel state},\ }\href
  {https://doi.org/10.1103/PhysRevLett.50.1153} {\bibfield  {journal} {\bibinfo
   {journal} {Phys. Rev. Lett.}\ }\textbf {\bibinfo {volume} {50}},\ \bibinfo
  {pages} {1153} (\bibinfo {year} {1983})}\BibitemShut {NoStop}%
\bibitem [{\citenamefont {Affleck}\ and\ \citenamefont
  {Haldane}(1987)}]{Affleck}%
  \BibitemOpen
  \bibfield  {author} {\bibinfo {author} {\bibfnamefont {I.}~\bibnamefont
  {Affleck}}\ and\ \bibinfo {author} {\bibfnamefont {F.~D.~M.}\ \bibnamefont
  {Haldane}},\ }\bibfield  {title} {\bibinfo {title} {Critical theory of
  quantum spin chains},\ }\href {https://doi.org/10.1103/PhysRevB.36.5291}
  {\bibfield  {journal} {\bibinfo  {journal} {Phys. Rev. B}\ }\textbf {\bibinfo
  {volume} {36}},\ \bibinfo {pages} {5291} (\bibinfo {year}
  {1987})}\BibitemShut {NoStop}%
\bibitem [{\citenamefont {den Nijs}\ and\ \citenamefont
  {Rommelse}(1989)}]{denNijs}%
  \BibitemOpen
  \bibfield  {author} {\bibinfo {author} {\bibfnamefont {M.}~\bibnamefont {den
  Nijs}}\ and\ \bibinfo {author} {\bibfnamefont {K.}~\bibnamefont {Rommelse}},\
  }\bibfield  {title} {\bibinfo {title} {Preroughening transitions in crystal
  surfaces and valence-bond phases in quantum spin chains},\ }\href
  {https://doi.org/10.1103/PhysRevB.40.4709} {\bibfield  {journal} {\bibinfo
  {journal} {Phys. Rev. B}\ }\textbf {\bibinfo {volume} {40}},\ \bibinfo
  {pages} {4709} (\bibinfo {year} {1989})}\BibitemShut {NoStop}%
\bibitem [{\citenamefont {Anfuso}\ and\ \citenamefont {Rosch}(2007)}]{Anfuso}%
  \BibitemOpen
  \bibfield  {author} {\bibinfo {author} {\bibfnamefont {F.}~\bibnamefont
  {Anfuso}}\ and\ \bibinfo {author} {\bibfnamefont {A.}~\bibnamefont {Rosch}},\
  }\bibfield  {title} {\bibinfo {title} {Fragility of string orders},\ }\href
  {https://doi.org/10.1103/PhysRevB.76.085124} {\bibfield  {journal} {\bibinfo
  {journal} {Phys. Rev. B}\ }\textbf {\bibinfo {volume} {76}},\ \bibinfo
  {pages} {085124} (\bibinfo {year} {2007})}\BibitemShut {NoStop}%
\bibitem [{\citenamefont {Wierschem}\ and\ \citenamefont
  {Sengupta}(2014)}]{WS3}%
  \BibitemOpen
  \bibfield  {author} {\bibinfo {author} {\bibfnamefont {K.}~\bibnamefont
  {Wierschem}}\ and\ \bibinfo {author} {\bibfnamefont {P.}~\bibnamefont
  {Sengupta}},\ }\bibfield  {title} {\bibinfo {title} {Quenching the haldane
  gap in spin-1 heisenberg antiferromagnets},\ }\href
  {https://doi.org/10.1103/PhysRevLett.112.247203} {\bibfield  {journal}
  {\bibinfo  {journal} {Phys. Rev. Lett.}\ }\textbf {\bibinfo {volume} {112}},\
  \bibinfo {pages} {247203} (\bibinfo {year} {2014})}\BibitemShut {NoStop}%
\bibitem [{\citenamefont {Pollmann}\ \emph {et~al.}(2012)\citenamefont
  {Pollmann}, \citenamefont {Berg}, \citenamefont {Turner},\ and\ \citenamefont
  {Oshikawa}}]{Pollmann1}%
  \BibitemOpen
  \bibfield  {author} {\bibinfo {author} {\bibfnamefont {F.}~\bibnamefont
  {Pollmann}}, \bibinfo {author} {\bibfnamefont {E.}~\bibnamefont {Berg}},
  \bibinfo {author} {\bibfnamefont {A.~M.}\ \bibnamefont {Turner}},\ and\
  \bibinfo {author} {\bibfnamefont {M.}~\bibnamefont {Oshikawa}},\ }\bibfield
  {title} {\bibinfo {title} {Symmetry protection of topological phases in
  one-dimensional quantum spin systems},\ }\href
  {https://doi.org/10.1103/PhysRevB.85.075125} {\bibfield  {journal} {\bibinfo
  {journal} {Phys. Rev. B}\ }\textbf {\bibinfo {volume} {85}},\ \bibinfo
  {pages} {075125} (\bibinfo {year} {2012})}\BibitemShut {NoStop}%
\bibitem [{\citenamefont {Moukouri}\ and\ \citenamefont
  {Eidelstein}(2012)}]{ME}%
  \BibitemOpen
  \bibfield  {author} {\bibinfo {author} {\bibfnamefont {S.}~\bibnamefont
  {Moukouri}}\ and\ \bibinfo {author} {\bibfnamefont {E.}~\bibnamefont
  {Eidelstein}},\ }\bibfield  {title} {\bibinfo {title} {Universality class of
  the mott transition in two dimensions},\ }\href
  {https://doi.org/10.1103/PhysRevB.86.155112} {\bibfield  {journal} {\bibinfo
  {journal} {Phys. Rev. B}\ }\textbf {\bibinfo {volume} {86}},\ \bibinfo
  {pages} {155112} (\bibinfo {year} {2012})}\BibitemShut {NoStop}%
\bibitem [{\citenamefont {Sakai}\ and\ \citenamefont
  {Takahashi}(1989)}]{Sakai}%
  \BibitemOpen
  \bibfield  {author} {\bibinfo {author} {\bibfnamefont {T.}~\bibnamefont
  {Sakai}}\ and\ \bibinfo {author} {\bibfnamefont {M.}~\bibnamefont
  {Takahashi}},\ }\bibfield  {title} {\bibinfo {title} {The ground state of
  quasi-one-dimensional heisenberg antiferromagnets},\ }\href@noop {}
  {\bibfield  {journal} {\bibinfo  {journal} {Journal of the Physical Society
  of Japan}\ }\textbf {\bibinfo {volume} {58}},\ \bibinfo {pages} {3131}
  (\bibinfo {year} {1989})}\BibitemShut {NoStop}%
\bibitem [{\citenamefont {Koga}\ and\ \citenamefont {Kawakami}(2000)}]{Koga}%
  \BibitemOpen
  \bibfield  {author} {\bibinfo {author} {\bibfnamefont {A.}~\bibnamefont
  {Koga}}\ and\ \bibinfo {author} {\bibfnamefont {N.}~\bibnamefont
  {Kawakami}},\ }\bibfield  {title} {\bibinfo {title} {Quantum phase
  transitions for the haldane system in higher dimensions: A mixed-spin cluster
  expansion approach},\ }\href {https://doi.org/10.1103/PhysRevB.61.6133}
  {\bibfield  {journal} {\bibinfo  {journal} {Phys. Rev. B}\ }\textbf {\bibinfo
  {volume} {61}},\ \bibinfo {pages} {6133} (\bibinfo {year}
  {2000})}\BibitemShut {NoStop}%
\bibitem [{\citenamefont {Kim}\ and\ \citenamefont {Birgeneau}(2000)}]{Kim}%
  \BibitemOpen
  \bibfield  {author} {\bibinfo {author} {\bibfnamefont {Y.~J.}\ \bibnamefont
  {Kim}}\ and\ \bibinfo {author} {\bibfnamefont {R.~J.}\ \bibnamefont
  {Birgeneau}},\ }\bibfield  {title} {\bibinfo {title} {Monte carlo study of
  the $s=\frac{1}{2}$ and $s=1$ heisenberg antiferromagnet on a spatially
  anisotropic square lattice},\ }\href
  {https://doi.org/10.1103/PhysRevB.62.6378} {\bibfield  {journal} {\bibinfo
  {journal} {Phys. Rev. B}\ }\textbf {\bibinfo {volume} {62}},\ \bibinfo
  {pages} {6378} (\bibinfo {year} {2000})}\BibitemShut {NoStop}%
\bibitem [{\citenamefont {Matsumoto}\ \emph {et~al.}(2001)\citenamefont
  {Matsumoto}, \citenamefont {Yasuda}, \citenamefont {Todo},\ and\
  \citenamefont {Takayama}}]{Matsumoto2001}%
  \BibitemOpen
  \bibfield  {author} {\bibinfo {author} {\bibfnamefont {M.}~\bibnamefont
  {Matsumoto}}, \bibinfo {author} {\bibfnamefont {C.}~\bibnamefont {Yasuda}},
  \bibinfo {author} {\bibfnamefont {S.}~\bibnamefont {Todo}},\ and\ \bibinfo
  {author} {\bibfnamefont {H.}~\bibnamefont {Takayama}},\ }\bibfield  {title}
  {\bibinfo {title} {Ground-state phase diagram of quantum heisenberg
  antiferromagnets on the anisotropic dimerized square lattice},\ }\href
  {https://doi.org/10.1103/PhysRevB.65.014407} {\bibfield  {journal} {\bibinfo
  {journal} {Phys. Rev. B}\ }\textbf {\bibinfo {volume} {65}},\ \bibinfo
  {pages} {014407} (\bibinfo {year} {2001})}\BibitemShut {NoStop}%
\bibitem [{\citenamefont {Wierschem}\ and\ \citenamefont
  {Sengupta}(2012)}]{WS1}%
  \BibitemOpen
  \bibfield  {author} {\bibinfo {author} {\bibfnamefont {K.}~\bibnamefont
  {Wierschem}}\ and\ \bibinfo {author} {\bibfnamefont {P.}~\bibnamefont
  {Sengupta}},\ }\bibfield  {title} {\bibinfo {title} {Dimensional crossover in
  spin-1 heisenberg antiferromagnets: a quantum monte carlo study},\ }\href
  {https://doi.org/10.1088/1742-6596/400/3/032112} {\bibfield  {journal}
  {\bibinfo  {journal} {Journal of Physics: Conference Series}\ }\textbf
  {\bibinfo {volume} {400}},\ \bibinfo {pages} {032112} (\bibinfo {year}
  {2012})}\BibitemShut {NoStop}%
\bibitem [{\citenamefont {Sandvik}(2010)}]{SSE}%
  \BibitemOpen
  \bibfield  {author} {\bibinfo {author} {\bibfnamefont {A.~W.}\ \bibnamefont
  {Sandvik}},\ }\bibfield  {title} {\bibinfo {title} {Computational studies of
  quantum spin systems},\ }\href {https://doi.org/10.1063/1.3518900} {\bibfield
   {journal} {\bibinfo  {journal} {AIP Conference Proceedings}\ }\textbf
  {\bibinfo {volume} {1297}},\ \bibinfo {pages} {135} (\bibinfo {year}
  {2010})}\BibitemShut {NoStop}%
\bibitem [{\citenamefont {Sylju\aa{}sen}\ and\ \citenamefont
  {Sandvik}(2002)}]{directedloop}%
  \BibitemOpen
  \bibfield  {author} {\bibinfo {author} {\bibfnamefont {O.~F.}\ \bibnamefont
  {Sylju\aa{}sen}}\ and\ \bibinfo {author} {\bibfnamefont {A.~W.}\ \bibnamefont
  {Sandvik}},\ }\bibfield  {title} {\bibinfo {title} {Quantum monte carlo with
  directed loops},\ }\href {https://doi.org/10.1103/PhysRevE.66.046701}
  {\bibfield  {journal} {\bibinfo  {journal} {Phys. Rev. E}\ }\textbf {\bibinfo
  {volume} {66}},\ \bibinfo {pages} {046701} (\bibinfo {year}
  {2002})}\BibitemShut {NoStop}%
\bibitem [{\citenamefont {Cheng}\ \emph {et~al.}(2016)\citenamefont {Cheng},
  \citenamefont {Zaletel}, \citenamefont {Barkeshli}, \citenamefont
  {Vishwanath},\ and\ \citenamefont {Bonderson}}]{3DSPT}%
  \BibitemOpen
  \bibfield  {author} {\bibinfo {author} {\bibfnamefont {M.}~\bibnamefont
  {Cheng}}, \bibinfo {author} {\bibfnamefont {M.}~\bibnamefont {Zaletel}},
  \bibinfo {author} {\bibfnamefont {M.}~\bibnamefont {Barkeshli}}, \bibinfo
  {author} {\bibfnamefont {A.}~\bibnamefont {Vishwanath}},\ and\ \bibinfo
  {author} {\bibfnamefont {P.}~\bibnamefont {Bonderson}},\ }\bibfield  {title}
  {\bibinfo {title} {Translational symmetry and microscopic constraints on
  symmetry-enriched topological phases: A view from the surface},\ }\href
  {https://doi.org/10.1103/PhysRevX.6.041068} {\bibfield  {journal} {\bibinfo
  {journal} {Phys. Rev. X}\ }\textbf {\bibinfo {volume} {6}},\ \bibinfo {pages}
  {041068} (\bibinfo {year} {2016})}\BibitemShut {NoStop}%
\bibitem [{\citenamefont {Evertz}(2003)}]{Evertz1}%
  \BibitemOpen
  \bibfield  {author} {\bibinfo {author} {\bibfnamefont {H.~G.}\ \bibnamefont
  {Evertz}},\ }\bibfield  {title} {\bibinfo {title} {The loop algorithm},\
  }\href {https://doi.org/10.1080/0001873021000049195} {\bibfield  {journal}
  {\bibinfo  {journal} {Advances in Physics}\ }\textbf {\bibinfo {volume}
  {52}},\ \bibinfo {pages} {1} (\bibinfo {year} {2003})},\ \Eprint
  {https://arxiv.org/abs/https://doi.org/10.1080/0001873021000049195}
  {https://doi.org/10.1080/0001873021000049195} \BibitemShut {NoStop}%
\bibitem [{\citenamefont {Affleck}\ \emph {et~al.}(1989)\citenamefont
  {Affleck}, \citenamefont {Gepner}, \citenamefont {Schulz},\ and\
  \citenamefont {Ziman}}]{AGSZ}%
  \BibitemOpen
  \bibfield  {author} {\bibinfo {author} {\bibfnamefont {I.}~\bibnamefont
  {Affleck}}, \bibinfo {author} {\bibfnamefont {D.}~\bibnamefont {Gepner}},
  \bibinfo {author} {\bibfnamefont {H.}~\bibnamefont {Schulz}},\ and\ \bibinfo
  {author} {\bibfnamefont {T.}~\bibnamefont {Ziman}},\ }\bibfield  {title}
  {\bibinfo {title} {Critical behaviour of spin-s heisenberg antiferromagnetic
  chains: analytic and numerical results},\ }\href@noop {} {\bibfield
  {journal} {\bibinfo  {journal} {Journal of Physics A: Mathematical and
  General}\ }\textbf {\bibinfo {volume} {22}},\ \bibinfo {pages} {511}
  (\bibinfo {year} {1989})}\BibitemShut {NoStop}%
\bibitem [{\citenamefont {Singh}\ \emph {et~al.}(1989)\citenamefont {Singh},
  \citenamefont {Fisher},\ and\ \citenamefont {Shankar}}]{SFS}%
  \BibitemOpen
  \bibfield  {author} {\bibinfo {author} {\bibfnamefont {R.~R.~P.}\
  \bibnamefont {Singh}}, \bibinfo {author} {\bibfnamefont {M.~E.}\ \bibnamefont
  {Fisher}},\ and\ \bibinfo {author} {\bibfnamefont {R.}~\bibnamefont
  {Shankar}},\ }\bibfield  {title} {\bibinfo {title} {Spin-1/2
  antiferromagnetic xxz chain: New results and insights},\ }\href
  {https://doi.org/10.1103/PhysRevB.39.2562} {\bibfield  {journal} {\bibinfo
  {journal} {Phys. Rev. B}\ }\textbf {\bibinfo {volume} {39}},\ \bibinfo
  {pages} {2562} (\bibinfo {year} {1989})}\BibitemShut {NoStop}%
\bibitem [{\citenamefont {Giamarchi}\ and\ \citenamefont {Schulz}(1989)}]{GS}%
  \BibitemOpen
  \bibfield  {author} {\bibinfo {author} {\bibfnamefont {T.}~\bibnamefont
  {Giamarchi}}\ and\ \bibinfo {author} {\bibfnamefont {H.~J.}\ \bibnamefont
  {Schulz}},\ }\bibfield  {title} {\bibinfo {title} {Correlation functions of
  one-dimensional quantum systems},\ }\href
  {https://doi.org/10.1103/PhysRevB.39.4620} {\bibfield  {journal} {\bibinfo
  {journal} {Phys. Rev. B}\ }\textbf {\bibinfo {volume} {39}},\ \bibinfo
  {pages} {4620} (\bibinfo {year} {1989})}\BibitemShut {NoStop}%
\bibitem [{\citenamefont {Barber}(1983)}]{Barber}%
  \BibitemOpen
  \bibfield  {author} {\bibinfo {author} {\bibfnamefont {M.}~\bibnamefont
  {Barber}},\ }\href
  {https://www.amazon.com/Phase-Transitions-Critical-Phenomena-8/dp/0122203089}
  {\emph {\bibinfo {title} {Phase Transitions and Critical Phenomena}}},\
  edited by\ \bibinfo {editor} {\bibfnamefont {C.}~\bibnamefont {Domb}}\ and\
  \bibinfo {editor} {\bibfnamefont {J.~L.}\ \bibnamefont {Lebowitz}},\
  Vol.~\bibinfo {volume} {8}\ (\bibinfo  {publisher} {Academic Press},\
  \bibinfo {address} {London},\ \bibinfo {year} {1983})\BibitemShut {NoStop}%
\bibitem [{\citenamefont {Lubensky}\ and\ \citenamefont
  {Rubin}(1975)}]{Lubensky}%
  \BibitemOpen
  \bibfield  {author} {\bibinfo {author} {\bibfnamefont {T.~C.}\ \bibnamefont
  {Lubensky}}\ and\ \bibinfo {author} {\bibfnamefont {M.~H.}\ \bibnamefont
  {Rubin}},\ }\bibfield  {title} {\bibinfo {title} {Critical phenomena in
  semi-infinite systems. i. $\ensuremath{\epsilon}$ expansion for positive
  extrapolation length},\ }\href {https://doi.org/10.1103/PhysRevB.11.4533}
  {\bibfield  {journal} {\bibinfo  {journal} {Phys. Rev. B}\ }\textbf {\bibinfo
  {volume} {11}},\ \bibinfo {pages} {4533} (\bibinfo {year}
  {1975})}\BibitemShut {NoStop}%
\bibitem [{\citenamefont {Campostrini}\ \emph {et~al.}(2002)\citenamefont
  {Campostrini}, \citenamefont {Hasenbusch}, \citenamefont {Pelissetto},
  \citenamefont {Rossi},\ and\ \citenamefont {Vicari}}]{CHPRV}%
  \BibitemOpen
  \bibfield  {author} {\bibinfo {author} {\bibfnamefont {M.}~\bibnamefont
  {Campostrini}}, \bibinfo {author} {\bibfnamefont {M.}~\bibnamefont
  {Hasenbusch}}, \bibinfo {author} {\bibfnamefont {A.}~\bibnamefont
  {Pelissetto}}, \bibinfo {author} {\bibfnamefont {P.}~\bibnamefont {Rossi}},\
  and\ \bibinfo {author} {\bibfnamefont {E.}~\bibnamefont {Vicari}},\
  }\bibfield  {title} {\bibinfo {title} {Critical exponents and equation of
  state of the three-dimensional heisenberg universality class},\ }\href
  {https://doi.org/10.1103/PhysRevB.65.144520} {\bibfield  {journal} {\bibinfo
  {journal} {Phys. Rev. B}\ }\textbf {\bibinfo {volume} {65}},\ \bibinfo
  {pages} {144520} (\bibinfo {year} {2002})}\BibitemShut {NoStop}%
\bibitem [{\citenamefont {Deng}\ \emph {et~al.}(2005)\citenamefont {Deng},
  \citenamefont {Bl\"ote},\ and\ \citenamefont {Nightingale}}]{Deng}%
  \BibitemOpen
  \bibfield  {author} {\bibinfo {author} {\bibfnamefont {Y.}~\bibnamefont
  {Deng}}, \bibinfo {author} {\bibfnamefont {H.~W.~J.}\ \bibnamefont
  {Bl\"ote}},\ and\ \bibinfo {author} {\bibfnamefont {M.~P.}\ \bibnamefont
  {Nightingale}},\ }\bibfield  {title} {\bibinfo {title} {Surface and bulk
  transitions in three-dimensional $\mathrm{O}(n)$ models},\ }\href
  {https://doi.org/10.1103/PhysRevE.72.016128} {\bibfield  {journal} {\bibinfo
  {journal} {Phys. Rev. E}\ }\textbf {\bibinfo {volume} {72}},\ \bibinfo
  {pages} {016128} (\bibinfo {year} {2005})}\BibitemShut {NoStop}%
\bibitem [{\citenamefont {Diehl}\ and\ \citenamefont {Dietrich}(1980)}]{DD1}%
  \BibitemOpen
  \bibfield  {author} {\bibinfo {author} {\bibfnamefont {H.}~\bibnamefont
  {Diehl}}\ and\ \bibinfo {author} {\bibfnamefont {S.}~\bibnamefont
  {Dietrich}},\ }\bibfield  {title} {\bibinfo {title} {Scaling laws and surface
  exponents from renormalization group equations},\ }\href
  {https://doi.org/https://doi.org/10.1016/0375-9601(80)90783-5} {\bibfield
  {journal} {\bibinfo  {journal} {Physics Letters A}\ }\textbf {\bibinfo
  {volume} {80}},\ \bibinfo {pages} {408 } (\bibinfo {year}
  {1980})}\BibitemShut {NoStop}%
\bibitem [{\citenamefont {Diehl}\ and\ \citenamefont {N\"usser}(1986)}]{DN}%
  \BibitemOpen
  \bibfield  {author} {\bibinfo {author} {\bibfnamefont {H.~W.}\ \bibnamefont
  {Diehl}}\ and\ \bibinfo {author} {\bibfnamefont {A.}~\bibnamefont
  {N\"usser}},\ }\bibfield  {title} {\bibinfo {title} {Critical behavior of the
  nonlinear \ensuremath{\sigma} model with a free surface: The ``ordinary''
  transition in 2+\ensuremath{\epsilon} dimensions},\ }\href
  {https://doi.org/10.1103/PhysRevLett.56.2834} {\bibfield  {journal} {\bibinfo
   {journal} {Phys. Rev. Lett.}\ }\textbf {\bibinfo {volume} {56}},\ \bibinfo
  {pages} {2834} (\bibinfo {year} {1986})}\BibitemShut {NoStop}%
\bibitem [{\citenamefont {Diehl}\ and\ \citenamefont {Shpot}(1994)}]{DS1}%
  \BibitemOpen
  \bibfield  {author} {\bibinfo {author} {\bibfnamefont {H.~W.}\ \bibnamefont
  {Diehl}}\ and\ \bibinfo {author} {\bibfnamefont {M.}~\bibnamefont {Shpot}},\
  }\bibfield  {title} {\bibinfo {title} {Surface critical behavior in fixed
  dimensions $d<4$: Nonanalyticity of critical surface enhancement and massive
  field theory approach},\ }\href {https://doi.org/10.1103/PhysRevLett.73.3431}
  {\bibfield  {journal} {\bibinfo  {journal} {Phys. Rev. Lett.}\ }\textbf
  {\bibinfo {volume} {73}},\ \bibinfo {pages} {3431} (\bibinfo {year}
  {1994})}\BibitemShut {NoStop}%
\bibitem [{\citenamefont {Diehl}\ and\ \citenamefont {Shpot}(1998)}]{DS2}%
  \BibitemOpen
  \bibfield  {author} {\bibinfo {author} {\bibfnamefont {H.~W.}\ \bibnamefont
  {Diehl}}\ and\ \bibinfo {author} {\bibfnamefont {M.}~\bibnamefont {Shpot}},\
  }\bibfield  {title} {\bibinfo {title} {Massive field-theory approach to
  surface critical behavior in three-dimensional systems},\ }\href
  {https://doi.org/10.1016/S0550-3213(98)00489-1} {\bibfield  {journal}
  {\bibinfo  {journal} {Nuclear Physics B}\ }\textbf {\bibinfo {volume}
  {528}},\ \bibinfo {pages} {595} (\bibinfo {year} {1998})}\BibitemShut
  {NoStop}%
\bibitem [{\citenamefont {Gliozzi}\ \emph {et~al.}(2015)\citenamefont
  {Gliozzi}, \citenamefont {Liendo}, \citenamefont {Meineri},\ and\
  \citenamefont {Rago}}]{GLMR}%
  \BibitemOpen
  \bibfield  {author} {\bibinfo {author} {\bibfnamefont {F.}~\bibnamefont
  {Gliozzi}}, \bibinfo {author} {\bibfnamefont {P.}~\bibnamefont {Liendo}},
  \bibinfo {author} {\bibfnamefont {M.}~\bibnamefont {Meineri}},\ and\ \bibinfo
  {author} {\bibfnamefont {A.}~\bibnamefont {Rago}},\ }\bibfield  {title}
  {\bibinfo {title} {Boundary and interface cfts from the conformal
  bootstrap},\ }\href {https://doi.org/10.1007/JHEP05(2015)036} {\bibfield
  {journal} {\bibinfo  {journal} {Journal of High Energy Physics}\ }\textbf
  {\bibinfo {volume} {2015}},\ \bibinfo {pages} {36} (\bibinfo {year}
  {2015})}\BibitemShut {NoStop}%
\bibitem [{\citenamefont {Diehl}\ and\ \citenamefont {Dietrich}(1981)}]{DD2}%
  \BibitemOpen
  \bibfield  {author} {\bibinfo {author} {\bibfnamefont {H.~W.}\ \bibnamefont
  {Diehl}}\ and\ \bibinfo {author} {\bibfnamefont {S.}~\bibnamefont
  {Dietrich}},\ }\bibfield  {title} {\bibinfo {title} {Field-theoretical
  approach to multicritical behavior near free surfaces},\ }\href
  {https://doi.org/10.1103/PhysRevB.24.2878} {\bibfield  {journal} {\bibinfo
  {journal} {Phys. Rev. B}\ }\textbf {\bibinfo {volume} {24}},\ \bibinfo
  {pages} {2878} (\bibinfo {year} {1981})}\BibitemShut {NoStop}%
\end{thebibliography}%

\end{document}